\documentclass[aps,prd,          groupedaddress,amsmath,amssymb]{revtex4}
\usepackage{slashed}
\usepackage{epsfig}

\begin{document}

\def\Re  {{\rm Re}}
\def\Im  {{\rm Im}}
\def\KFS#1{K^{{#1}}_{\scriptscriptstyle\rm F\!.S\!.}}
\def\Ret {\widetilde{{\rm Re}}}
\def\veps{\varepsilon}
\def\hc    {{\rm h.c.}}
\renewcommand{\eqref}[1]{Eq.~(\ref{#1})}
\newcommand{\fact}{{\mathrm{C}}}
\newcommand{\figref}[1]{Fig.~\ref{#1}}
\newcommand{\tabref}[1]{Tab.~\ref{#1}}
\newcommand{\rA}{{\mathrm{A}}}
\newcommand{\rB}{{\mathrm{B}}}
\newcommand{\rR}{{\mathrm{R}}}
\newcommand{\rV}{{\mathrm{V}}}
\def\bom#1{{\mbox{\boldmath $#1$}}}
\def\ket#1{|{#1}\rangle}
\def\bra#1{\langle{#1}|}
\def\sd{\tilde{d}}
\def\sq{\tilde{q}}
\def\su{\tilde{u}}
\def\mq{m_q}
\def\mqi{m_{q}}
\def\mqj{m_{q}}
\def\mqk{m_{q}}
\def\mql{m_{q}}
\def\msqi{m_{\tilde{q}_i}}
\def\msqj{m_{\tilde{q}_j}}
\def\msqk{m_{\tilde{q}_k}}
\def\msql{m_{\tilde{q}_l}}
\def\mgl{m_{\tilde{g}}}

\def\ca{\tilde{\chi}^\pm_1}
\def\cb{\tilde{\chi}^\pm_2}
\def\na{\tilde{\chi}^0_1}
\def\nb{\tilde{\chi}^0_2}
\def\nc{\tilde{\chi}^0_3}
\def\nd{\tilde{\chi}^0_4}
\def\ncd{\tilde{\chi}^0_{3,4}}
\def\sq{\tilde{q}}

\def\cA{{\cal A}} \def\cB{{\cal B}} \def\cC{{\cal C}} \def\cD{{\cal D}}
\def\cE{{\cal E}} \def\cF{{\cal F}} \def\cG{{\cal G}} \def\cH{{\cal H}}
\def\cI{{\cal I}} \def\cJ{{\cal J}} \def\cK{{\cal K}} \def\cL{{\cal L}}
\def\cM{{\cal M}} \def\cN{{\cal N}} \def\cO{{\cal O}} \def\cP{{\cal P}}
\def\cQ{{\cal Q}} \def\cR{{\cal R}} \def\cS{{\cal S}} \def\cT{{\cal T}}
\def\cU{{\cal U}} \def\cV{{\cal V}} \def\cW{{\cal W}} \def\cX{{\cal X}}
\def\cY{{\cal Y}} \def\cZ{{\cal Z}} 

\def\d{{\rm d}}
\def\eps{\epsilon}
\def\MSbar{\overline{\rm MS}}

\def\lp{\left. }
\def\rp{\right. }
\def\lr{\left( }
\def\rr{\right) }
\def\le{\left[ }
\def\re{\right] }
\def\lg{\left\{ }
\def\rg{\right\} }
\def\lb{\left| }
\def\rb{\right| }

\def\bsp#1\esp{\begin{split}#1\end{split}}

\def\beq{\begin{equation}}
\def\eeq{\end{equation}}
\def\bea{\begin{eqnarray}}
\def\eea{\end{eqnarray}}

\preprint{IPHC-PHENO-10-02}
\preprint{LPSC 10-050}
\title{Threshold Resummation for Gaugino Pair Production at Hadron
 Colliders}
\author{Jonathan Debove$^a$}
\author{Benjamin Fuks$^b$}
\author{Michael Klasen$^a$}
\email[]{klasen@lpsc.in2p3.fr}
\affiliation{$^a$ Laboratoire de Physique Subatomique et de Cosmologie,
 Universit\'e Joseph Fourier/CNRS-IN2P3/INPG,
 53 Avenue des Martyrs, F-38026 Grenoble, France \\
 $^b$ Institut Pluridisciplinaire Hubert Curien/D\'epartement Recherche Subatomique,
 Universit\'e de Strasbourg/CNRS-IN2P3,
 23 Rue du Loess, F-67037 Strasbourg, France}
\date{\today}
\begin{abstract}
We present a complete analysis of threshold resummation effects on direct
light and heavy gaugino pair production at the Tevatron and the LHC. Based on a
new perturbative calculation at next-to-leading order of SUSY-QCD, which includes
also squark mixing effects, we resum soft gluon radiation in the threshold region
at leading and next-to-leading logarithmic accuracy, retaining at the same time
the full SUSY-QCD
corrections in the finite coefficient function. This allows us to correctly match
the resummed to the perturbative cross section. Universal subleading logarithms
are resummed in full matrix form. We find that threshold resummation
slightly increases and considerably stabilizes the invariant mass spectra and
total cross sections with respect to the next-to-leading order calculation. For
future reference, we present total cross sections and their theoretical errors
in tabular form for several commonly used SUSY benchmark points, gaugino pairs,
and hadron collider energies.
\end{abstract}
\maketitle


\vspace*{-92mm}
\noindent IPHC-PHENO-10-02\\
\noindent LPSC 10-050\\
\vspace*{80mm}

\section{Introduction}
\label{sec:1}

The Minimal Supersymmetric Standard Model (MSSM) \cite{Nilles:1983ge}
continues to be one of the best-motivated and most intensely studied
extensions of the Standard Model (SM) of particle physics. It introduces
a symmetry between fermionic and bosonic degrees of freedom in nature and
predicts one fermionic (bosonic) supersymmetric (SUSY) partner for each
bosonic (fermionic) SM particle. Consequently, it allows to stabilize the gap
between the electroweak and the Planck scale \cite{Witten:1981nf} and to
unify the three gauge couplings at energies of ${\cal O}(10^{16})$ GeV
\cite{Dimopoulos:1981yj}. It also contains a stable lightest supersymmetric
particle (LSP), which interacts only weakly and represents therefore an
excellent candidate for cold dark matter \cite{Ellis:1983ew}. Spin partners
of the SM particles have not yet been observed, and in order to remain a
viable solution to the hierarchy problem, SUSY must be broken at low energy
via soft terms in the Lagrangian. As a consequence, the SUSY particles
must be massive in comparison to their SM counterparts, and the Tevatron and
the LHC will perform a conclusive search covering a wide range of masses up
to the TeV scale.

The production of SUSY particles at hadron colliders has been studied at
leading order (LO) of perturbative QCD since the 1980s \cite{Barger:1983wc}.
More recently, previously neglected electroweak contributions
\cite{Bozzi:2005sy}, polarization effects \cite{Gehrmann:2004xu}, and the
violation of flavor \cite{Bozzi:2007me} and $CP$ symmetry \cite{Alan:2007rp}
have been considered at this order. Next-to-leading order (NLO) corrections
have been computed within QCD since the late 1990s \cite{Beenakker:1996ch,%
Beenakker:1999xh} and recently also within the electroweak theory
\cite{Hollik:2007wf}. Resummation at the next-to-leading logarithmic (NLL)
level has been achieved in the small transverse-momentum region for sleptons
and gauginos \cite{Bozzi:2006fw}, in the threshold region for sleptons,
squarks and gluinos \cite{Bozzi:2007qr}, and for sleptons and additional
neutral gauge bosons also in both regions simultaneously \cite{Bozzi:2007tea}.

As the fermionic partners of the electroweak gauge and Higgs bosons, the
four neutralino $(\tilde{\chi}^0_i$, $i=1,\dots,4$) and chargino
$(\tilde{\chi}^\pm_i$, $i=1,2$) mass eigenstates are superpositions of the
neutral and charged gaugino and higgsino interaction eigenstates. Their
decays into leptons and missing transverse energy, carried away by the $\na$
LSP, are easily identifiable at hadron colliders. The lighter mass
eigenstates are accessible not only at the LHC with center-of-mass energies
$\sqrt{S}$ of 7 to 14 TeV, but also at Run II of the Tevatron ($\sqrt{S}=1.96$
TeV), where the production of $\ca\nb$ pairs decaying into trilepton
final states is a gold-plated SUSY discovery channel \cite{Aaltonen:2008pv}.
For this particular channel, threshold resummation has been studied previously
within the minimal supergravity (mSUGRA) model \cite{Li:2007ih}. It was
found that threshold resummation could increase the total cross section by up
to 4.7\% at the Tevatron relative to the NLO prediction, i.e.\ was
significant even relatively far from the hadronic production threshold.

In this work, we extend and improve on these published results in several
respects. First, we extend the original NLO calculation \cite{Beenakker:1999xh}
by also including squark mixing effects. Second, we include not only the 
QCD, but also the SUSY-QCD virtual loop contributions in the hard coefficient
function of the resummed cross section, which therefore reproduces, when
expanded, the correct NLO SUSY-QCD cross section in the threshold region.
Third, we resum not only the diagonal, but the full matrix contributions
coming from the anomalous dimension, thereby including all universal
subleading terms and full singlet mixing. However, we refrain from resumming
constant terms that are known to factorize and exponentiate in Drell-Yan like
processes, but not in more complex processes like gaugino pair production with
several interfering Born diagrams. For the Tevatron, we consider not only the
production of $\ca\nb$, but also of $\nb\nb$ and $\ca\ca$ pairs. In particular
the latter can have significantly larger cross sections than trilepton
production due to the $s$-channel exchange of massless photons. For the LHC,
we concentrate on predictions for its initial center-of-mass energy of
$\sqrt{S}=$7 TeV and include also the production of heavy gaugino ($\cb$,
$\ncd$) combinations, where threshold effects and direct gaugino pair
production (as opposed to the production from squark and gluino cascade
decays) will be more important. However, we will also show cross sections for
$\sqrt{S}=$14 TeV for comparison.

The remainder of this paper is organized as follows: In Sec.\ \ref{sec:2},
we present briefly our LO and NLO SUSY-QCD calculations, focusing on squark
mixing, the ultraviolet renormalization procedure and the dipole subtraction
method employed for the cancellation of infrared divergences among virtual and
real contributions. In Sec.\ \ref{sec:3}, we describe the threshold resummation
formalism in Mellin space, starting from the refactorization of the
perturbative cross section and the exponentiation of the eikonal function,
giving then the explicit form of the resummed logarithms at NLL order, and
finally describing various ways to improve on the original formalism. Sec.\
\ref{sec:4} contains numerical results for various gaugino pair production
cross sections at the Tevatron and at the LHC in both graphical and tabular
form. It also includes a comprehensive analysis of theoretical errors coming
from scale and parton density uncertainties. We summarize our results in Sec.\
\ref{sec:5}. Our conventions for the couplings of quarks and squarks to weak
gauge bosons and gauginos are defined in the Appendix. 

\section{Gaugino pair production at fixed order in perturbative QCD}
\label{sec:2}

\begin{figure}
 \centering
 \epsfig{file=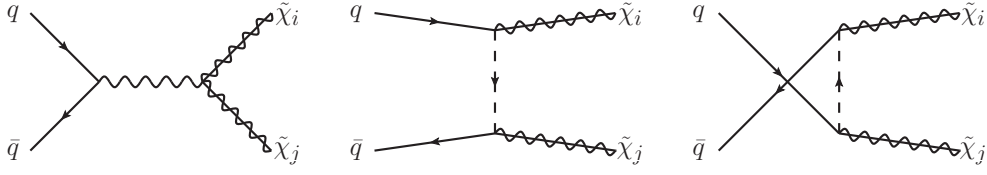,width=.75\columnwidth}
 \caption{\label{fig:1}Tree-level Feynman diagrams for the production of gaugino
          pairs.}
\end{figure}

At hadron colliders, two gauginos $\tilde{\chi}_{i,j}$ with masses
$m_{\tilde{\chi}_{i,j}}$ and four-momenta $p_{1,2}$ can be produced at LO of
perturbation theory only through the annihilation of quarks $q$ and antiquarks
$\bar{q}'$ with four-momenta $p_{a,b}$,
\bea
  q(p_a)\ \bar{q}'(p_b) &\to& \tilde{\chi}_i(p_1)\ \tilde{\chi}_j(p_2),
  \label{eq:1}
\eea
since gluons couple neither to electroweak gauge/Higgs bosons nor to their
SUSY partners \footnote{At the tree-level, we neglect $s$-channel Higgs-boson
exchanges, which may be produced by initial-state gluons through top-quark
loops.}. As shown in Fig.\ \ref{fig:1}, the production can proceed either
%
%
through the $s$-channel exchange of a photon ($\gamma$), neutral ($Z$) or
charged ($W$) weak gauge boson, depending on the charge of the final state,
or through the $t$- and $u$-channel exchange of a squark ($\tilde{q}$),
where the Mandelstam variables
\bea
 s~=~(p_a+p_b)^2 &,& t~=~(p_a-p_1)^2 \mbox{~~,~and~~} u~=~(p_a-p_2)^2
\eea
and their mass-subtracted counterparts
$t_{\tilde{\chi}_{i,j}}=t-m_{\tilde{\chi}_{i,j}}^2$ and
$u_{\tilde{\chi}_{i,j}}=u-m_{\tilde{\chi}_{i,j}}^2$ are defined in the usual
way.

\subsection{Leading order cross section}

\renewcommand{\arraystretch}{3}
\begin{table}
 \caption{\label{tab:1}$s$-channel charges for gaugino pair production. The
  masses of the $W$- and $Z$-bosons are $m_W$ and $m_Z$.\\ }
 \begin{tabular}{|c|cccc|}
\hline
  & $Q_L^{su}$ & $Q_L^{st}$ & $Q_R^{su}$ & $Q_R^{st}$\\
  \hline \hline 
   $\tilde{\chi}_i^0\tilde{\chi}_j^0$ & 
    $\frac{L_{q q^\prime Z} O^{\prime\prime L}_{ij}}{s - m^2_Z}$ & 
    $\frac{L_{q q^\prime Z} O^{\prime\prime R}_{ij}}{s - m^2_Z}$ & 
    $\frac{R_{q q^\prime Z} O^{\prime\prime R}_{ij}}{s - m^2_Z}$ & 
    $\frac{R_{q q^\prime Z} O^{\prime\prime L}_{ij}}{s - m^2_Z}$ \\
   $\tilde{\chi}_i^+\tilde{\chi}_j^0$ & 
    $\frac{L_{q q^\prime W}^\ast O^{L\ast}_{ji}}{s - m^2_W}$ & 
    $\frac{L_{q q^\prime W}^\ast O^{R\ast}_{ji}}{s - m^2_W}$ & 
    0 & 
    0\\
   $\tilde{\chi}_i^+\tilde{\chi}_j^-$ & 
    $-\frac{e_q x_W \delta_{q q^\prime} \delta_{ij}}{4 s} + \frac{L_{q q^\prime Z} O^{\prime L}_{ij}}{s - m^2_Z}~$  & 
    $-\frac{e_q x_W \delta_{q q^\prime} \delta_{ij}}{4 s} + \frac{L_{q q^\prime Z} O^{\prime R}_{ij}}{s - m^2_Z}~$  & 
    $-\frac{e_q x_W \delta_{q q^\prime} \delta_{ij}}{4 s} + \frac{R_{q q^\prime Z} O^{\prime R}_{ij}}{s - m^2_Z}~$  & 
    $-\frac{e_q x_W \delta_{q q^\prime} \delta_{ij}}{4 s} + \frac{R_{q q^\prime Z} O^{\prime L}_{ij}}{s - m^2_Z}$  \\
 \hline
 \end{tabular}
\end{table}
\renewcommand{\arraystretch}{1}
The LO differential cross section for given (anti-)quark helicities $h_{a,b}$
can be expressed in the compact form
\bea
  \frac{\d \sigma_{q\bar{q}'}^{h_a,h_b}}{\d t} &=& 
  \frac{4 \pi \alpha^2}{C_A (1+\delta_{\chi_i\chi_j}) x_W^2 s^2}
     \Big\{(1-h_a) (1+h_b) \Big[ 
       \left| Q^u_{LL} \right|^2 u_{\tilde{\chi}_i} u_{\tilde{\chi}_j} + 
       \left| Q_{LL}^t \right|^2 t_{\tilde{\chi}_i} t_{\tilde{\chi}_j} + 
       2 {\rm Re} [Q_{LL}^{u\ast} Q_{LL}^t] m_{\tilde{\chi}_i} m_{\tilde{\chi}_j} s \Big]\\
  &&\hspace*{30mm} + (1+h_a) (1-h_b) \Big[ 
       \left| Q_{RR}^u \right|^2 u_{\tilde{\chi}_i} u_{\tilde{\chi}_j} + 
       \left| Q_{RR}^t \right|^2 t_{\tilde{\chi}_i} t_{\tilde{\chi}_j} + 
       2 {\rm Re} [Q_{RR}^{u\ast} Q_{RR}^t] m_{\tilde{\chi}_i} m_{\tilde{\chi}_j} s \Big] \nonumber\\
  &&\hspace*{30mm} + (1+h_a) (1+h_b) \Big[ 
       \left| Q_{RL}^u \right|^2 u_{\tilde{\chi}_i} u_{\tilde{\chi}_j} + 
       \left| Q_{RL}^t \right|^2 t_{\tilde{\chi}_i} t_{\tilde{\chi}_j} - 
       2 {\rm Re} [Q_{RL}^{u\ast} Q_{RL}^t] (u t - m^2_{\tilde{\chi}_i} m^2_{\tilde{\chi}_j})
       \Big] \nonumber\\ 
  &&\hspace*{30mm} + (1-h_a) (1-h_b) \Big[ 
       \left| Q_{LR}^u \right|^2 u_{\tilde{\chi}_i} u_{\tilde{\chi}_j} + 
       \left| Q_{LR}^t \right|^2 t_{\tilde{\chi}_i} t_{\tilde{\chi}_j} - 
       2 {\rm Re} [Q_{LR}^{u\ast} Q_{LR}^t] (u t - m^2_{\tilde{\chi}_i} m^2_{\tilde{\chi}_j})
       \Big]
     \Big\}\nonumber
\eea
by employing the generalized charges
\bea
 Q_{XY}^t ~=~ 2 Q_X^{st} \delta_{XY} - \sum_{\sq} \frac{X^\ast_{\tilde{q} q \tilde{\chi}_i} Y_{\tilde{q} q^\prime \tilde{\chi}_j}}{t - m^2_{\tilde{q}}}
 &~~{\rm and}~~&
 Q_{XY}^u ~=~ 2 Q_X^{su} \delta_{XY} + \sum_{\sq} \frac{X^\ast_{\tilde{q} q \tilde{\chi}_j} Y_{\tilde{q} q^\prime \tilde{\chi}_i}}{u - m^2_{\tilde{q}}} \ ,
\eea
where the $s$-channel charges $Q_{X}^{st}$ and $Q_{X}^{su}$ are given in Tab.\
\ref{tab:1}, $m_{\tilde{q}}$ is the mass of the squark exchanged in the $t$- or
$u$-channel, $X,Y\in\{L,R\}$ relate to the left- and right-handed (s)quark
couplings to weak gauge bosons and gauginos listed in App.\ \ref{sec:a},
and the couplings of the latter among each other depend on the neutral and charged
gaugino-higgsino mixing matrices $N$, $U$ and $V$ through the bilinear
combinations \cite{Nilles:1983ge}
\bea
  O^L_{ij} = \frac{1}{2 \sqrt{2}} N_{i4} V^\ast_{j2} - \frac12 N_{i2} V^\ast_{j1} 
    & {\rm ~~~~and~~~~}& 
    O^R_{ij} = -\frac{1}{2 \sqrt{2}} N_{i3}^\ast U_{j2} -\frac 12 N_{i2}^\ast U_{j1} \ ,~\\
  O^{\prime L}_{ij} = \frac{1}{2 c_W} V_{i1} V_{j1}^\ast + \frac{1}{4 c_W} V_{i2} V_{j2}^\ast - \frac{1}{2 c_W}  \delta_{ij} x_W 
    &{\rm ~~~~and~~~~}& 
    O^{\prime R}_{ij} =  \frac{1}{2 c_W} U_{i1}^\ast U_{j1} +  \frac{1}{4 c_W} U_{i2}^\ast U_{j2} - \frac{1}{2 c_W} \delta_{ij} x_W \ ,~ \\ 
  O^{\prime\prime L}_{ij} = \frac{1}{4 c_W} \Big[N_{i3} N_{j3}^\ast - N_{i4}N_{j4}^\ast \Big]
    &{\rm ~~~~and~~~~}& 
  O^{\prime\prime R}_{ij} = \frac{1}{4 c_W} \Big[ - N_{i3}^\ast N_{j3} + N_{i4}^\ast N_{j4} \Big] \ .~
\eea 
$\alpha=e^2/(4\pi)$ is the electromagnetic fine structure constant, $x_W=1-c_W^2=
s_W^2=\sin^2\theta_W$ is the squared sine of the electroweak mixing angle, and
$C_A=3$, $C_F=4/3$ and $\alpha_s=g_s^2/(4\pi)$ are the QCD color factors
and coupling constant. Unpolarized cross sections are obtained by averaging the
polarized ones with
\bea
 \d\sigma_{q\bar{q}'}^{(0)}&=&
 \frac{\d\sigma^{ 1, 1}_{q\bar{q}'} + \d\sigma^{ 1,-1}_{q\bar{q}'} +
       \d\sigma^{-1, 1}_{q\bar{q}'} + \d\sigma^{-1,-1}_{q\bar{q}'}}{4}.
\eea

\subsection{Virtual corrections}

\begin{figure}
 \centering
 \epsfig{file=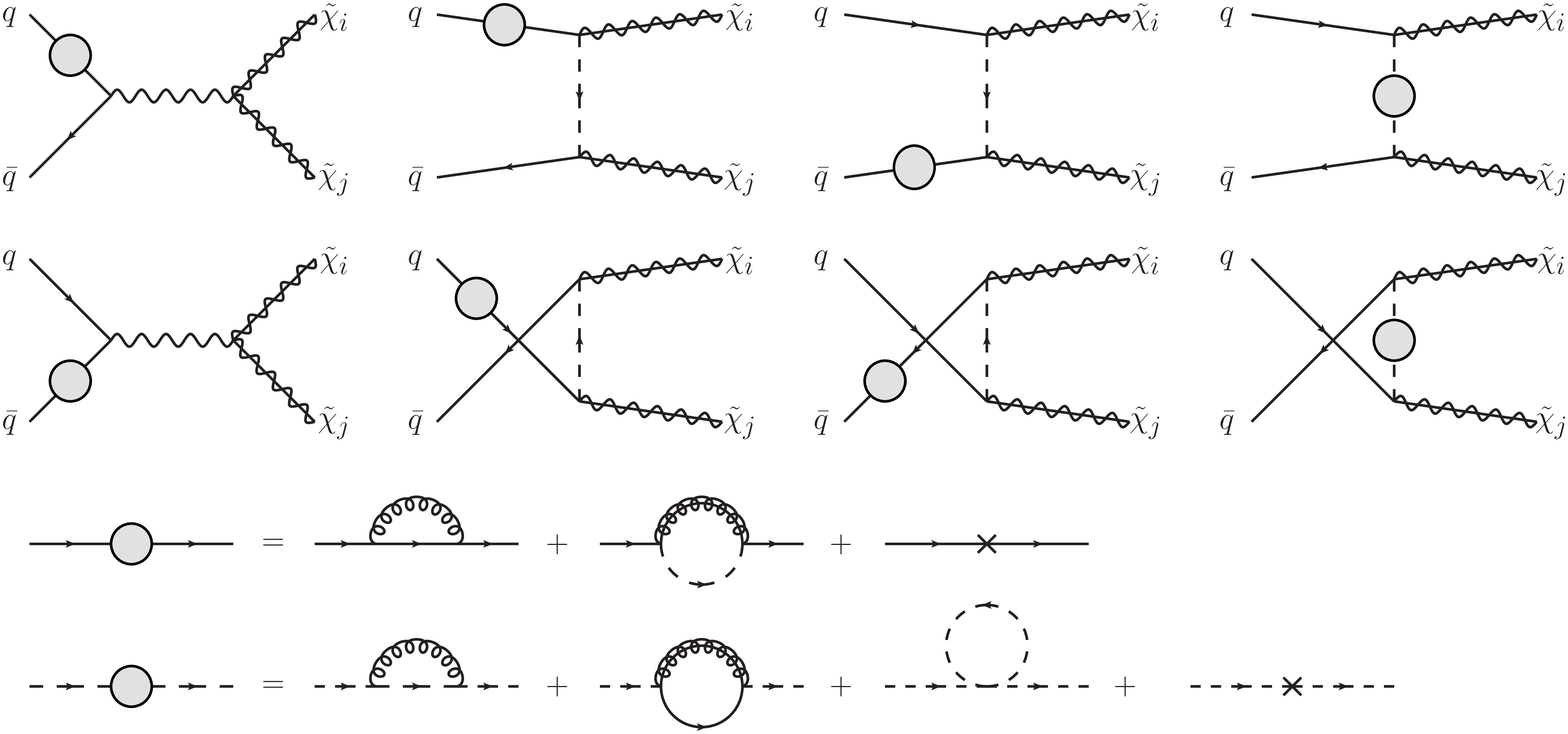,width=\columnwidth}
 \caption{\label{fig:2}Self-energy insertions (top) and contributions (bottom)
          to the production of gaugino pairs.}
\end{figure}

\begin{figure}
 \centering
 \epsfig{file=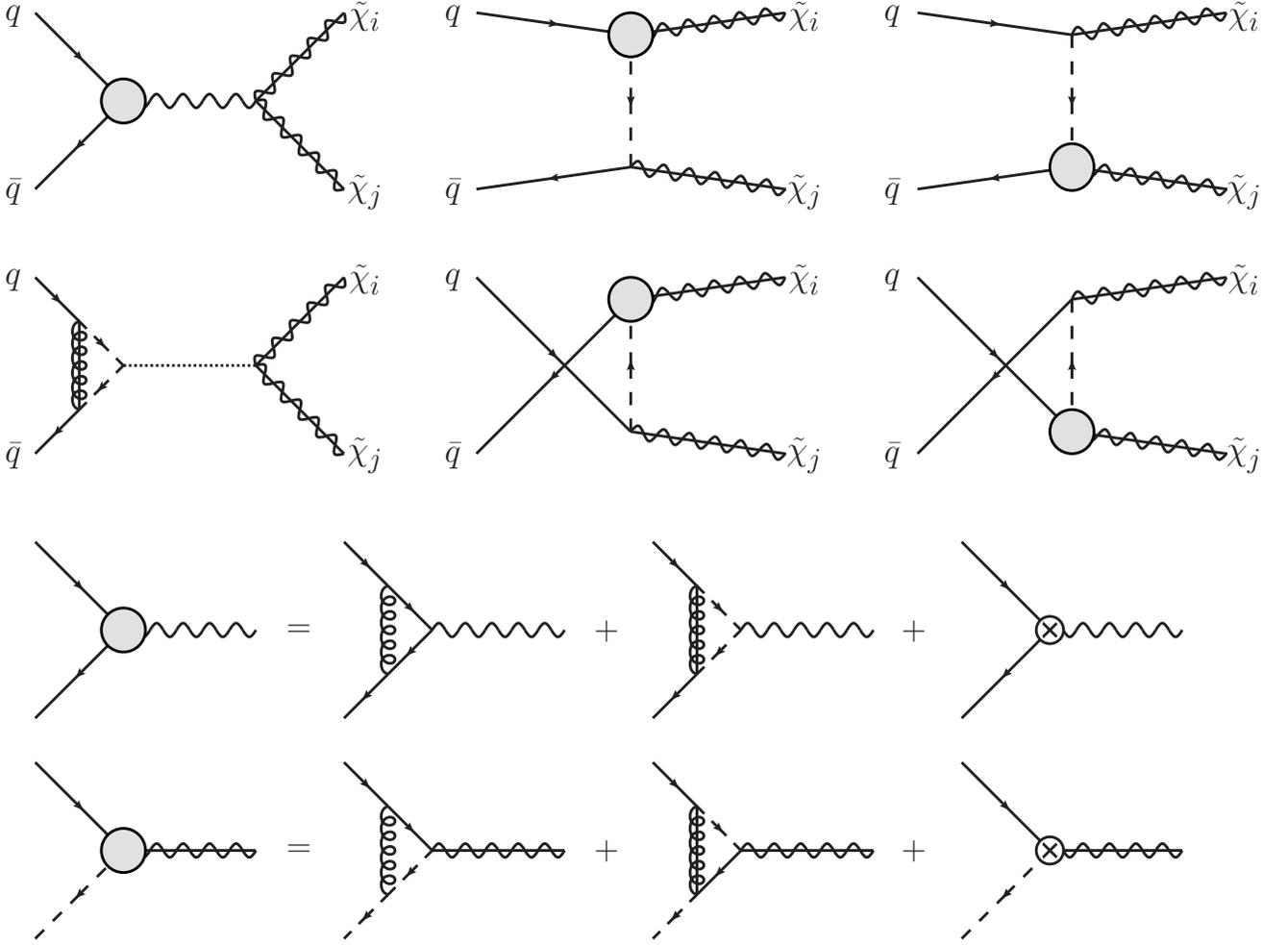,width=\columnwidth}
 \caption{\label{fig:3}Vertex correction insertions (top) and contributions
          (bottom) to the production of gaugino pairs.}
\end{figure}

\begin{figure}
 \centering
 \epsfig{file=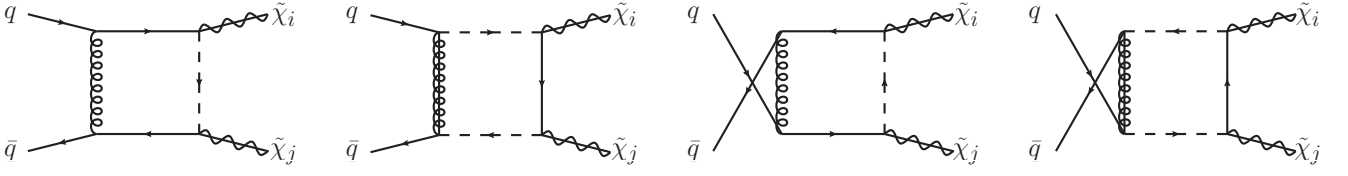,width=\columnwidth}
 \caption{\label{fig:4}Box diagrams contributing to the production of gaugino
          pairs at NLO.}
\end{figure}

At NLO of SUSY-QCD, ${\cal O}(\alpha_s)$, the cross section for gaugino pair
production receives contributions from the interference of the virtual one-loop
diagrams shown in Figs.\ \ref{fig:2}--\ref{fig:4} with the tree-level diagrams
shown in Fig.\
%
%
\ref{fig:1} on the one hand and from real gluon (Fig.\ \ref{fig:5}) and
(anti-)quark emission diagrams on the other hand, where the latter are obtained by
crossing the final-state gluon in Fig.\ \ref{fig:5} with the initial-state
antiquark (Fig.\ \ref{fig:6}) or quark (not shown).
All diagrams have been evaluated analytically with self-written FORM programs
and cross-checked independently with self-written MATHEMATICA programs.

The virtual self-energy diagrams for left- and right-handed quarks $q_{L,R}=
P_{L,R}\,q={1\over2}(1\mp\gamma_5) q$ (Fig.\ \ref{fig:2}, third line),
\bea
    \Sigma^{(g)}_{L,R}(p)&=&-\frac{g_s^2 C_F}{16\pi^2}\big[(D-2)
    \slashed{p}B_1(p,\mql,0)+D\mql B_0(p,\mql,0)\big]P_{L,R},\\
    \Sigma^{(\tilde{g})}_{L~~\,}(p)&=&-\frac{g_s^2 C_F}{8\pi^2}
    \sum_{i=1}^{2}\Big[\slashed{p}B_1(p,\mgl,\msqi)R_{i1}^{\sq\ast}R_{i1}^{\sq}
    +\mgl B_0(p,\mgl,\msqi)R_{i2}^{\sq\ast}R_{i1}^{\sq}\Big]P_L,~~{\rm and}\\
    \Sigma^{(\tilde{g})}_{R~~\,}(p)&=&-\frac{g_s^2 C_F}{8\pi^2}
    \sum_{i=1}^{2}\Big[\slashed{p}B_1(p,\mgl,\msqi)R_{i2}^{\sq\ast}R_{i2}^{\sq}
    +\mgl B_0(p,\mgl,\msqi)R_{i1}^{\sq\ast}R_{i2}^{\sq}\Big]P_R,
\eea
expanded as usual into vector (V) and scalar (S) parts $\Sigma(p)=[\Sigma^{V}_L
(p^2)\,\slashed{p}+\Sigma^{S}_L(p^2)]P_L+(L\leftrightarrow R)$, as well as
those for squarks (Fig.\ \ref{fig:2}, fourth line),
\bea
    \Sigma^{(g)}_{ij}(p^2)&=&-\frac{g_s^2 C_F}{16\pi^2}
    \Big[p^2\big(B_0(p,\msqi,0)-2B_1(p,\msqi,0)
      +B_{21}(p,\msqi,0)\big)+DB_{22}(p,\msqi,0)\Big]\delta_{ij},\label{eq:9}\\
    \Sigma^{(\tilde{g})}_{ij}(p^2)&=&-\frac{g_s^2 C_F}{4\pi^2}
                \Big\{\Big[p^2\big(B_1(p,\mq,\mgl)+B_{21}(p,\mq,\mgl)\big)
      +DB_{22}(p,\mq,\mgl)\Big]
    \delta_{ij}
      \nonumber\\
    &&\hspace*{12mm}-~\mq\mgl~ B_0(p,\mq,\mgl)
    \big(R_{i1}^{\sq}R_{j2}^{\sq\ast}+R_{i2}^{\sq}R_{j1}^{\sq\ast}\big)\Big\},~~{\rm and}\label{eq:10}\\
    \Sigma^{(\tilde{q})}_{ij}(p^2)&=&\frac{g_s^2 C_F}{16\pi^2}
    \sum_{k=1}^2S_{ki}^{\sq}S_{kj}^{\sq\ast} A_0(\msqk) ~~~{\rm with}~~~
    S_{ki}=R^{\sq\ast}_{k1}R^{\sq}_{i1}-R^{\sq\ast}_{k2}R^{\sq}_{i2},
    \label{eq:11}
\eea
contain ultraviolet (UV) divergences in the scalar integrals $B_{0,1,...}
(p,m_1,m_2)$ \cite{Passarino:1978jh}, which exhibit themselves as $1/\eps$
poles in $D=4-2\eps$ dimensions \footnote{As it is customary, we absorb a factor
of $(2\pi\mu_R)^{4-D}$ in the definition of the scalar integrals, where $\mu_R$ is
the renormalization scale.}. They must therefore be absorbed through a suitable
%
%
renormalization procedure into the fundamental wave functions, mass parameters,
and coupling constants of the SUSY-QCD Lagrangian
\bea
    \cL&=&\Big[\bar{q}^0_L i\slashed{\partial} q^0_L 
        - \bar{q}^0_R m^0_q q^0_L
        + (L\leftrightarrow R)\Big]
        + \Big[\sum_{i=1}^2(\partial_\mu\sq^0_i)^\dagger(\partial^\mu\sq^0_i)
        - \sq^{0\dagger}_i(m^2_{\sq})^0_{ii}\sq^0_i\Big]
        + \, ... \, .\label{eq:12}
\eea
The two components of the unrenormalized squark field $\sq^0$ correspond
originally to the left- and right-handed chiralities of the unrenormalized SM
quark field $q^0$, but mix due to the fact that soft SUSY-breaking and Higgs
terms render the $2\times2$-dimensional mass matrix $(m^2_{\sq})^0$ non-diagonal
\cite{Nilles:1983ge}. In Eq.\ (\ref{eq:12}) we have diagonalized this mass matrix
with the squark rotation matrix $R^{\sq\,0}$, so that the components $i=1\,(2)$
of the squark field correspond to the squark mass eigenvalues $m_{\sq_i}^0$. The
squark self-energies in Eqs.\ (\ref{eq:9})-(\ref{eq:11}) thus also carry indices
$i,j=1,2$ corresponding to the (outgoing and incoming) squark mass eigenstates.
Multiplicative renormalization is achieved perturbatively by expanding the
renormalization constants,
\bea
 q^0_{L,R}~=~ \lr 1        ~~+\frac{1}{2}\,\delta Z_{q}~~\rr q_{L,R} \,
 &{\rm and}&~
 ~m^0_q ~~~\,=~ ~m_q~~~+~\delta m_q~~~\,, \\
 \sq^0_i  ~~~\,=~ \lr \delta_{ij}+\frac{1}{2}\,\delta Z_{\sq,ij} \rr \sq_j ~~~
 &{\rm and}&
 (m^2_{\sq})^0_{ij}~=~ (m^2_{\sq})_{ij}+(\delta m^2_{\sq})_{ij},
\eea
with the usual factor of $1/2$ for the (s)quark wave functions. The renormalized
self-energies are then
\bea
    \hat{\Sigma}(p)&=&\Big[\Sigma^{V}_L(p^2)
    +\frac{1}{2}(\delta Z_q+\delta Z_q^\dagger)\Big]\slashed{p}P_L
    +\Big[\Sigma^{S}_L(p^2)
      -\frac{1}{2}(m_q\delta Z_q+\delta Z_q^\dagger m_q)-\delta m_q\Big]P_L
    +(L\leftrightarrow R)
\eea
for quarks and
\bea
    \hat{\Sigma}_{ij}(p^2)&=&\Sigma_{ij}(p^2)
    +\frac{1}{2}(\delta Z_{\sq,ij}+\delta Z^\ast_{\sq,ji})\,p^2
    -\frac{1}{2}\sum_{k=1}^2\le(m_{\sq}^2)_{ik}\delta Z_{\sq,kj}+
                               \delta Z^\ast_{\sq,ki}(m_{\sq}^2)_{kj}\re
    -(\delta m_{\sq}^2)_{ij}
\eea
for squarks.

We choose to renormalize the wave functions in the $\MSbar$-scheme, so that
the definition of the quark fields corresponds to the one employed in the
parton densities in the external hadrons \footnote{This is at variance with
Ref.\ \cite{Li:2007ih}, where the quark wave functions are renormalized on-shell,
thereby absorbing finite SUSY contributions into the definition of the quark
fields.}. In this scheme, the quark wave function counterterm
\bea
  \delta Z_q~=~\delta Z_q^{(g)}+\delta Z_q^{(\tilde{g})} ~&{\rm with}&~
  \delta Z_q^{(g)}~=~\delta Z_q^{(\tilde{g})}~=~-\frac{g_s^2 C_F}{16\pi^2}\Delta
  ~~{\rm and}~~ \Delta~=~{1\over\eps}-\gamma_E+\ln4\pi,
 \label{eq:qwfc}
\eea
defined as the UV-divergent plus universal finite parts of the on-shell
counterterm $-\Sigma^V_{L,R}(m_q^2)-m_q^2[\Sigma^{V\prime}_{L,R}(m_q^2)+
\Sigma^{V\prime}_{R,L}(m_q^2)]-m_q[\Sigma^{S\prime}_{L,R}(m_q^2)+
\Sigma^{S\prime}_{R,L}(m_q^2)] $ \cite{Bohm:2001yx}, is hermitian
($\delta Z_q=\delta Z_q^\dagger$) and the
same for left- and right-handed quarks. The superscripts $g$ and $\tilde{g}$ label
the gluon and gluino exchange contributions, respectively, and $\gamma_E$ is the
Euler constant. The squark wave function counterterms
\bea
 \delta Z_{\tilde{q},ij}&=&
 \delta Z_{\tilde{q},ij}^{(g)}+
 \delta Z_{\tilde{q},ij}^{(\tilde{g})}+
 \delta Z_{\tilde{q},ij}^{(\tilde{q})},
\eea
with
\bea
  \delta Z_{\tilde{q},ii}^{(g)}~=~-\delta Z_{\tilde{q},ii}^{(\tilde{g})}~=~
  \frac{g_s^2 C_F}{8\pi^2}\Delta ~&{\rm and}&~
  \delta Z_{\tilde{q},ii}^{(\tilde{q})}~=~0
  \label{eq:17}
\eea
for $i=j$ and
\bea
  \delta Z^{(g)}_{\tilde{q},ij}&=&0,\\
  \delta Z_{\tilde{q},ij}^{(\tilde{g})}&=&\frac{g_s^2 C_F}{4\pi^2}\frac{2\Delta}{\msqi^2
  -\msqj^2}\Big[\mq\mgl\big(R^{\sq}_{i1}R_{j2}^{\sq\ast}+R^{\sq}_{i2}R_{j1}^{\sq\ast}\big)\Big],~~{\rm and}\label{eq:19}\\
  \delta Z_{\tilde{q},ij}^{(\tilde{q})}&=&\frac{g_s^2 C_F}{16\pi^2}\frac{2\Delta}
  {\msqi^2-\msqj^2}\sum_{k=1}^{2}\msqk^2S_{ki}S_{kj}^\ast\label{eq:20}
\eea
for $i\neq j$, defined similarly as the UV-divergent plus universal finite parts
of the on-shell counterterms $-\widetilde{\rm Re}\,\Sigma^\prime_{ii}(\msqi^2)$
for $i=j$ and $2\,\widetilde{\rm Re}\,\Sigma_{ij}(\msqj^2)/(\msqi^2-\msqj^2)$ for
$i\neq j$ \cite{Bohm:2001yx}, enter only through the renormalization of the squark
mixing matrix,
\bea
  R^{\sq\,0}=R^{\sq}+\delta R^{\sq} ~&{\rm with}&~
  \delta R_{ij}^{\sq}~=~\frac{1}{4}\sum_{k=1}^2(\delta Z_{\tilde{q},ik}-
  \delta Z^\ast_{\tilde{q},ki})R_{kj}^{\sq}~=~
  \frac{1}{2}\sum_{k=1}^2\delta Z_{\tilde{q},ik}R_{kj}^{\sq},
  \label{eq:21}
\eea
since in the $\MSbar$-scheme the gluon and gluino contributions for $i=j$ in Eq.\
(\ref{eq:17}) cancel each other. In the last step of Eq.\ (\ref{eq:21}), we have
made use of the fact that in the $\MSbar$-scheme the squark wave-function
renormalization constants are anti-hermitian matrices ($\delta Z_{\tilde{q},ij}=
-\delta Z_{\tilde{q},ji}^{\ast}$). 
%
%

%
%
The (s)quark masses are renormalized in the on-shell scheme to make them
correspond to the physical masses. The quark mass counterterm is then defined
by $m_q\Sigma^V(m_q^2)+\Sigma^S(m_q^2)$ \cite{Bohm:2001yx} with the result
\bea
 \delta m_q&=&\delta m_q^{(g)}+\delta m_q^{(\tilde{g})}
\eea
and
\bea
  \delta m_q^{(g)}&=&-\frac{g_s^2 C_F}{16\pi^2} m_q \Big[(D-2) B_1(m_q,
  m_q,0)+DB_0(m_q,m_q,0)\Big],\\
  \delta m_q^{(\tilde{g})}&=&-\frac{g_s^2 C_F}{16\pi^2}\sum_{i=1}^{2}
  \Big[m_q B_1(m_q,\mgl,\msqi)
  +2\mgl B_0(m_q,\mgl,\msqi)\Re\big(R^{\sq\ast}_{i2}R^{\sq}_{i1}\big)\Big].
\eea
For our numerical results, we will set the masses of external quarks to zero
in accordance with the collinear factorization of quarks in hadrons.
The squark mass counterterm is defined by $\widetilde{\rm Re}[\Sigma_{ii}
(\msqi^2)]$. 
The result is
\bea
 \delta \msqi^{2}&=&\delta \msqi^{2(g)}+\delta\msqi^{2(\tilde{g})}+
\delta\msqi^{2(\tilde{q})}
\eea
with
\bea
  \delta \msqi^{2(g)}&=&\frac{g_s^2C_F}{8\pi^2}\msqi^2\Big[B_1(\msqi,\msqi,0)
  -B_0(\msqi,\msqi, 0)\Big],\\
  \delta\msqi^{2(\tilde{g})}&=&-\frac{g_s^2C_F}
  {4\pi^2}\Big[\msqi^2B_1(\msqi,\mq,\mgl)+\mq^2B_0(\msqi,\mq,\mgl)
  +A_0(\mgl)\nonumber \\
  &&\hspace*{13mm}-\,2\mq\mgl B_0(\msqi,\mq,\mgl)\,\Re(R_{i1}^{\sq}R_{i2}^{\sq\ast})\Big],~~{\rm and}\\
  \delta\msqi^{2(\tilde{q})}&=&\frac{g_s^2 C_F}{16\pi^2}\sum_{j=1}^2
  |S_{ji}|^2 A_0(\msqj).
\eea

Supersymmetric Ward identities link the quark-quark-gauge boson and
quark-squark-gaugino vertices to the weak gauge-boson and gaugino self-energies.
As the latter do not receive strong corrections at NLO, the former require no
further renormalization beyond the one for the (s)quark wave functions discussed
above. However, the artificial breaking of supersymmetry by the mismatch of two
gaugino and $(D-2)$ transverse vector degrees of freedom must be compensated
by a finite counterterm $\hat{g}=g[1-\alpha_sC_F/(8\pi)]$, effectively shifting
the quark-squark-gaugino scalar coupling constant $\hat{g}$ with respect to the
weak gauge coupling constant $g$ \cite{Beenakker:1996ch,AguilarSaavedra:2005pw}.

\subsection{Real corrections}

\begin{figure}
 \centering
 \epsfig{file=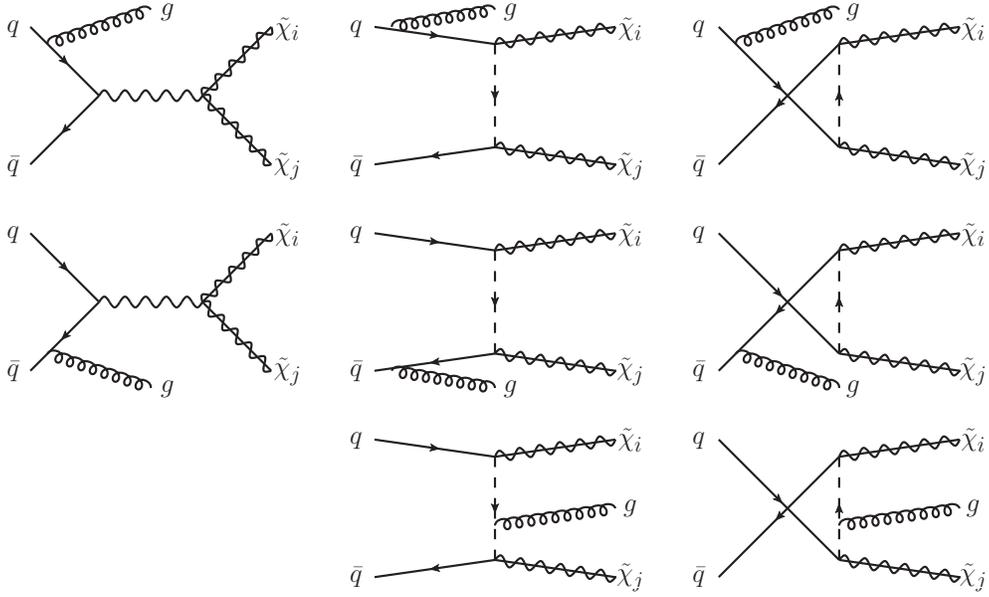,width=.75\columnwidth}
 \caption{\label{fig:5}Gluon emission diagrams contributing to the production
          of gaugino pairs at NLO.}
\end{figure}

\begin{figure}
 \centering
 \epsfig{file=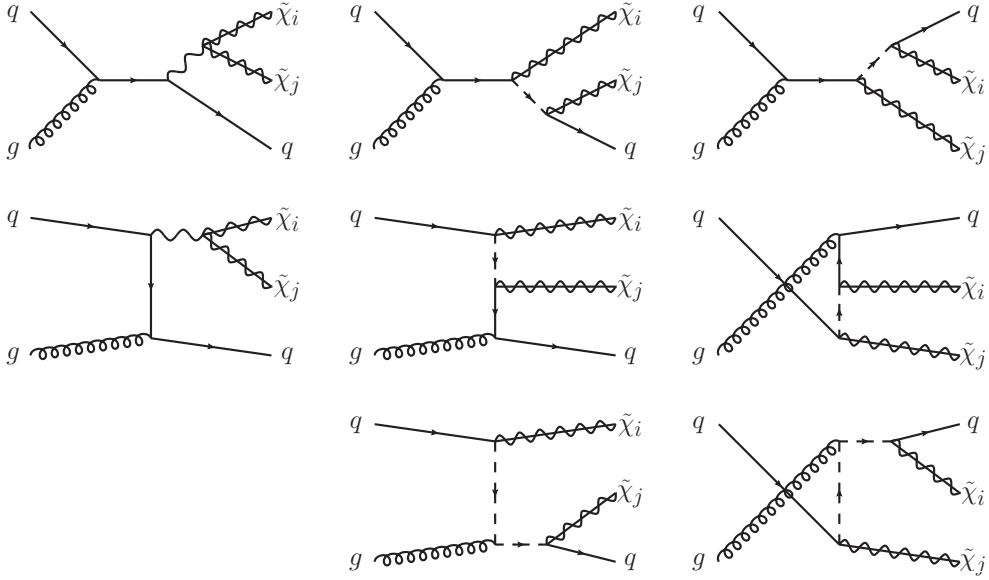,width=.75\columnwidth}
 \caption{\label{fig:6}Quark emission diagrams contributing to the production
          of gaugino pairs at NLO.}
\end{figure}

Apart from the (now UV-finite) virtual corrections $\d\sigma^{(V)}_{ab}$ to the LO
cross section $\d\sigma^{(0)}_{ab}$ described above, the NLO cross section
\bea
 \label{eq:34}
 \d\sigma^{(1)}_{ab}(p_a,p_b) &=&
 \int_{2+1} \left[ \left( \d\sigma^{(R)}_{ab}(p_a,p_b) \right)_{\eps=0} -
 \left( \sum_{\mathrm{dipoles}} \;\d\sigma^{(0)}_{ab}(p_a,p_b) \otimes 
 \;\d V_{\mathrm{dipole}}
 \right)_{\eps=0} \;\right]  \nonumber\\
 &+&  \int_2 \left[ \d\sigma^{(V)}_{ab}(p_a,p_b) + \d\sigma^{(0)}_{ab}(p_a,p_b) \otimes {\bom I}
 \right]_{\eps=0} \nonumber\\
 &+& \sum_{a'}\int_0^1 \d x \; \int_2 \left[ \d\sigma^{(0)}_{a'b}(xp_a,p_b) \otimes
     \left( {\bom P}+{\bom K} \right)^{a,a'}(x)
  +                                        \d\sigma^{(0)}_{aa'}(p_a,xp_b) \otimes
     \left( {\bom P}+{\bom K} \right)^{b,a'}(x) \right]_{\eps=0}
\eea
also receives contributions $\d\sigma^{(R)}_{ab}$ from real gluon (Fig.\
\ref{fig:5}), quark (Fig.\ \ref{fig:6}) and antiquark (not shown) emission
%
%
%
%
diagrams, where the emitted parton carries four-momentum $p_3$. In the
Catani-Seymour dipole formalism \cite{Catani:1996vz}, the real contributions are
rendered infrared (IR) finite by subtracting from them their soft and collinear
limits ($p_{a,b} \cdot p_3 \to 0$)
\bea
 \d\sigma^{(0)}_{ab}(p_a,p_b) \otimes \d V_{\mathrm{dipole}}&=& \sum_i
 \left[ {\cal D}^{a3,b}(p_1,p_2,p_{3};p_a,p_b) \;
 F_J^{(2)}({\widetilde p}_1, {\widetilde p}_{2};{\widetilde p}_{a3},p_b)
 + ( a \leftrightarrow b ) \right]
\eea
before integration over the three-particle final-state phase space. They can then
be evaluated in four dimensions (i.e.\ with $\eps=0$). In the case at hand of
two initial-state partons and no colored final state particles at LO, the only
dipole contribution comes from an initial state emitter and an initial state
spectator, e.g.\
\bea
 {\cal D}^{a3,b}(p_1,p_2,p_{3};p_a,p_b) &=&
 - \frac{1}{2 p_a \cdot p_3}  \;\frac{1}{x_{3,ab}}
 {}_{2,ab}\bra{{\widetilde 1},{\widetilde {2}};{\widetilde {a3}},b}
 \,\frac{{\bom T}_b \cdot {\bom T}_{a3}}{{\bom T}_{a3}^2} \;
 {\bom V}^{a3,b} \,
 \ket{{\widetilde 1},{\widetilde {2}};{\widetilde {a3}},b}_{2,ab}
\eea
(see Eq.\ (5.136) of Ref.\ \cite{Catani:1996vz}). The color charges ${\bom T}_
{a3,b}$ and splitting functions ${\bom V}^{a3,b}$ (Eqs.\ (5.145)-(5.148) of
Ref.\ \cite{Catani:1996vz}) act on Born-like squared matrix elements, which are
written here in terms of vectors $\ket{{\widetilde 1},{\widetilde {2}};
{\widetilde {a3}},b}_{2,ab}$ in color and helicity space. These matrix elements
involve an initial-state parton ${\widetilde {a3}}$ with momentum parallel to
$p_a$,
\bea
{\widetilde p}_{a3}^\mu ~=~ x_{3,ab} \,p_a^\mu  &,~~{\rm where}~~&
 x_{3,ab} ~=~ \frac{p_a\cdot p_b-p_3\cdot p_a-p_3\cdot p_b}{p_a\cdot p_b},
\eea
and rescaled four-momenta of the final-state gauginos
\beq
{\widetilde p_{1,2}}^{\mu} = p_{1,2}^{\mu} - \frac{2 p_{1,2} \cdot (K+{\widetilde K})}
{(K+{\widetilde K})^2} \;(K+{\widetilde K})^{\mu} + \frac{2 p_{1,2} \cdot K}
{K^2} \;{\widetilde K}^{\mu},
\eeq
where $K^{\mu} = p_a^\mu + p_b^\mu - p_3^\mu$ and ${\widetilde K}^{\mu}=
{\widetilde p_{a3}}^\mu + p_b^\mu$. The phase space function $F_J^{(2)}(
{\widetilde p}_1, {\widetilde p}_{2};{\widetilde p}_{a3},p_b)$ tends to zero with
$p_a \cdot p_3$ and ensures therefore that the LO cross section is IR-finite. To
compensate for the subtracted auxiliary dipole term $\d\sigma^{(0)}_{ab}(p_a,
p_b) \otimes \d V_{\mathrm{dipole}}$, the latter must be integrated analytically
over the full phase space of the emitted parton,
\beq
{\bom I} = \sum_{\mathrm{dipoles}} \;\int_1 \;\d V_{\mathrm{dipole}} \;\;,
\eeq
and added to the virtual cross section. The integrated dipole term is defined
explicitly in Eq.\ (10.15) of Ref.\ \cite{Catani:1996vz}; it contains all the
simple and double poles in $\eps$ necessary to cancel the IR singularities in
$\d\sigma^{(V)}_{ab}$. The insertion operators 
\bea
{\bom P}^{a,a'}(p_1,...,p_m,p_b;xp_a,x;\mu_F^2)&=&
\frac{\alpha_s}{2\pi} \;P^{aa'}(x) \;\frac{1}{{\bom T}_{a'}^2}
\left[
\sum_{i} {\bom T}_i \cdot {\bom T}_{a'}
\;\ln \frac{\mu_F^2}{2 xp_a \cdot p_i}
+ {\bom T}_b \cdot {\bom T}_{a'}
\;\ln \frac{\mu_F^2}{2 xp_a \cdot p_b} \right]
\label{eq:paa}
\eea
are directly related to the regularized Altarelli-Parisi splitting distributions
at ${\cal O}(\alpha_s)$,
\bea
 P^{qq}(x) &=& ~C_F \le \frac{1 + x^2}{(1-x)_+} +{3\over2}\delta(1-x)\re,
 \label{eq:pqq} \\
 P^{qg}(x) &=& ~C_F \le \frac{1 + (1-x)^2}{x}\re, \\
 P^{gq}(x) &=& ~T_R \;\le x^2 + (1-x)^2 \right],~~{\rm and} \\
 P^{gg}(x) &=& 2C_A  \le \left( \frac{1}{1-x} \right)_+ + \frac{1-x}{x}
 -1+x(1-x)\right] + \beta_0\,\delta(1-x), \label{eq:pgg}
\eea
where $\beta_0=11C_A/6-2N_fT_R/3$ and $\beta_1=(17C_A^2-5C_AN_f-3C_FN_f)/6$ are
the one- and two-loop coefficients of the QCD beta-function, $C_F=4/3$, $T_R=1/2$,
$C_A=3$, and $N_f$ is the number of quark flavors. They cancel the dependence of
the hadronic cross section on the factorization scale $\mu_F$ up to NLO accuracy.
The insertion operators
\beq
{\bom K}^{a,a'}(x)
= \frac{\alpha_s}{2\pi}
\left\{ \frac{}{} {\overline K}^{aa'}(x) - \KFS{aa'}(x)
 + \;\delta^{aa'} \; \sum_{i} {\bom T}_i \cdot {\bom T}_a
\;\frac{\gamma_i^{(1)}}{{\bom T}_i^2} \left[
\left( \frac{1}{1-x} \right)_+ + \delta(1-x) \right]
\right\} -
\frac{\alpha_s}{2\pi} {\bom T}_b \cdot {\bom T}_{a'} \frac{1}{{\bom T}_{a'}^2}
{\widetilde K}^{aa'}(x)
\label{eq:kaa}
\eeq
with $\gamma_{q}^{(1)}=3C_F/2$ and $\gamma_{g}^{(1)} = \beta_0$,
\bea
\label{okqq}
{\overline K}^{qq}(x) \;\,=\;\, {\overline K}^{{\bar q}{\bar q}}(x)
&=& C_F \left[
\left( \frac{2}{1-x} \ln\frac{1-x}{x} \right)_+
- (1+x) \ln\frac{1-x}{x} + (1-x) \right] 
- \delta(1-x) \left( 5 - \pi^2 \right) C_F\;\;, \\
{\overline K}^{qg}(x) \;\,= \;\,{\overline K}^{{\bar q}g}(x)
&=& P^{qg}(x) \ln\frac{1-x}{x} + C_F \;x \;\;, \\
\label{okgq}
{\overline K}^{gq}(x) \;\,=\;\, {\overline K}^{g{\bar q}}(x)
&=& P^{gq}(x) \ln\frac{1-x}{x} + T_R \;2x(1-x) \;\;, \\
{\overline K}^{gg}(x) &=& 2 C_A \left[
\left( \frac{1}{1-x} \ln\frac{1-x}{x} \right)_+
+ \left( \frac{1-x}{x} - 1 + x(1-x) \right) \ln\frac{1-x}{x} \right] \nonumber\\
&-& \delta(1-x) \left[ \left( \frac{50}{9} -\pi^2 \right) C_A -
\frac{16}{9} T_R N_f \right]
\;\;,\\
{\overline K}^{{\bar q}q}(x) \;\,=\;\, {\overline K}^{q{\bar q}}(x)
&=& 0,
\eea
and
\bea
{\widetilde K}^{ab}(x) &=& P^{ab}_{{\rm reg}}(x) \;\ln(1-x)
+ \;\delta^{ab} \,{\bom T}_a^2 \left[ \left( \frac{2}{1-x} \ln (1-x)
\right)_+ - \frac{\pi^2}{3} \delta(1-x) \right]
\eea
depend on the factorization scheme through the term $\KFS{aa'}(x)$, which vanishes
in the $\MSbar$-scheme, and also on the regular parts of the Altarelli-Parisi
splitting
%
\begin{figure}
 \centering
 \epsfig{file=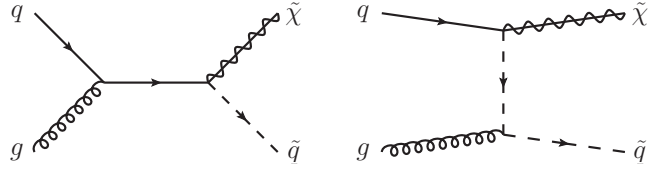,width=.5\columnwidth}
 \caption{\label{fig:7}Associated production of a gaugino and a virtual squark,
          decaying subsequently into a gaugino and a quark.}
\end{figure}
%
distributions given by $P^{ab}_{{\rm reg}}(x)=P^{ab}(x)$, if $a \neq b$, and
otherwise by
\bea
 P^{qq}_{{\rm reg}}(x) ~=~ - C_F \,(1 + x)~~&{\rm and}&~~
 P^{gg}_{{\rm reg}}(x) ~=~ 2\, C_A \left[ \frac{1-x}{x}  - 1 + x(1-x) \right].
\eea
The last line in Eq.\ (\ref{eq:34}) contains therefore the finite remainders that
are left after the factorization of collinear initial-state singularities into the
parton densities in the $\MSbar$-scheme at the factorization scale
$\mu_F$. As guaranteed by the Kinoshita-Lee-Nauenberg and factorization theorems,
the total NLO cross section is then not only UV-, but also IR-finite.

\begin{figure}
 \centering
 \epsfig{file=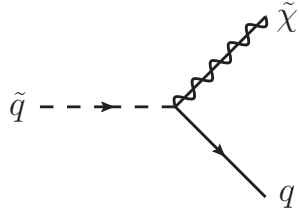,width=.25\columnwidth}
 \caption{\label{fig:8}Tree-level diagram for a squark decaying into a gaugino
          and a quark.}
\end{figure}

Finally, one subtlety must still be addressed: in Fig.\ \ref{fig:6}, the center
and right diagrams of lines one and three proceed through a squark propagator,
which can become on-shell if $m_{\tilde{q}}\geq m_{\tilde{\chi}}$ and $s\geq
(m_{\tilde{q}}+m_{\tilde{\chi}} )^2$. To avoid double counting, the resonance
contribution
\bea
 \d\sigma^{(\tilde{q})}_{qg}&=&\d\sigma(qg\to\tilde{\chi}\tilde{q}) ~
 {\rm BR}(\tilde{q}\to
 \tilde{\chi}q)
\eea
must be subtracted from the gaugino pair production process using the narrow-width
approximation, as it is identified experimentally as the associated production of
a gaugino and a squark (Fig.\ \ref{fig:7}), followed by the decay of the squark
into a gaugino and a quark (Fig.\ \ref{fig:8}).

\section{Threshold Resummation at Next-to-Leading Logarithmic Accuracy}
\label{sec:3}

In the previous section, we have demonstrated that all soft and collinear (IR)
singularities in the partonic NLO cross section $\d\sigma^{(1)}_{ab}$ either
cancel among virtual and real corrections or can be absorbed at the factorization
scale $\mu_F$ into parton density functions (PDFs) $f_{a,b/A,B}(x_{a,b},\mu_F^2)$,
which represent probability distributions for initial partons $a,b$ with
longitudinal momentum fraction $x_{a,b}$ in the external hadrons $A,B$ (e.g.\
protons or anti-protons). The QCD factorization theorem guarantees that the
observable hadronic cross section 
\bea
  M^2\frac{\d\sigma_{AB}}{\d M^2}(\tau)&=&\sum_{ab}
  \int_0^1 \!\d x_a \d x_b \d z[x_a f_{a/A}(x_a,\mu^2)] [x_b f_{b/B}(x_b,\mu^2)]
  \,[z\,\d\sigma_{ab}(z,M^2,\mu^2)]\,\delta(\tau-x_ax_bz)
  \label{eq:HadFacX}
\eea
can be obtained by convolving the process-dependent partonic cross section
\bea
 \d\sigma_{ab}(z, M^2, \mu^2)&=&
 \sum_{n=0}^\infty a_s^{n}(\mu^2)\,\d\sigma_{ab}^{(n)}(z, M^2, \mu^2)
  \label{eq:HarSca}
\eea
with universal PDFs. For the sake of simplicity, we will identify in the remainder
of this paper the factorization scale $\mu_F$ and the renormalization scale
%
%
$\mu_R$, which enters the partonic cross section through the perturbative
expansion in the (reduced) strong coupling constant $a_s(\mu_R^2)=\alpha_s
(\mu_R^2)/(2\pi)$ and also explicitly beyond LO, with the common scale $\mu$.
$M^2$ represents the invariant mass squared of the
(colorless) gaugino pair produced at LO, which carries the fractions $\tau=M^2/S$
and $z=M^2/s$ of the hadronic and partonic center-of-mass energies $S$ and $s=x_a
x_bS$, respectively. At LO, where no additional partons are produced besides the
gaugino pair, the partonic cross section
\beq
  \d\sigma_{ab}^{(0)}(z,M^2, \mu^2)=
    \sigma_{ab}^{(0)}(M^2)\,\delta(1-z)
  \label{eq:Sig0}
\eeq
is independent of $\mu$ and
has its support entirely at the point $z=1$. At higher order in QCD, the
cancellation of soft
and collinear parton emission among virtual and real corrections is restricted by
the phase space boundary of the latter. This leads to logarithmic contributions
$a_s^n [\ln^m(1-z)/(1-z)]_+$ with $m \leq 2n-1$ (see Eqs.\ (\ref{eq:paa}) and
(\ref{eq:kaa})), which become large close to the partonic threshold at $z\to1$.
Consequently, they spoil the convergence of the perturbative series and must be
resummed to all orders in $a_s$.

By applying a Mellin transform
\bea
  F(N)&=&\int_0^1 \d y \,y^{N-1} F(y)
  \label{eq:MelDef}
\eea
to the quantities $F=\sigma_{AB}$, $\sigma_{ab}$, and $f_{a,b/A,B}$ with $y=\tau$,
$z$, and $x_{a,b}$ in Eq.\ (\ref{eq:HadFacX}), the hadronic cross section can be
written as a simple product
\bea
  M^2\frac{\d\sigma_{AB}}{\d M^2}(N-1)&=&\sum_{ab} f_{a/A}(N,\mu^2) 
  f_{b/B}(N,\mu^2) \sigma_{ab}(N,M^2,\mu^2),
  \label{eq:HadFacN}
\eea
the large logarithms in $z\to1$ turn into large logarithms of the Mellin variable
$N$,
\beq
  \bigg(\frac{\ln^m(1-z)}{1-z}\bigg)_+ \longrightarrow \ln^{m+1}N~~{\rm etc.},
  \label{eq:TR:Log}
\eeq
and it is thus possible, in the $N\rightarrow\infty$ limit, to retain only the
leading powers in $N$. In particular, one can neglect among the $N$-moments of
the Altarelli-Parisi splitting functions at ${\cal O}(a_s)$
\bea
 P^{qq}(N)~=~\gamma_{qq}^{(1)}(N)
&=&C_F\bigg[\frac{3}{2}+\frac{1}{N(N+1)}-2\sum_{k=1}^N\frac{1}{k}\bigg]
        ~\to~C_F\bigg(\frac{3}{2}-2\ln\bar{N}\bigg)+\cO\lr\frac{1}{N}\rr,
 \label{eq:pqqn}\\
 P^{qg}(N)~=~\gamma_{qg}^{(1)}(N)
&=&C_F\bigg[\frac{2+N+N^2}{N(N^2-1)}\bigg]
        ~\to~\frac{C_F}{N},\\
 P^{gq}(N)~=~\gamma_{gq}^{(1)}(N)
&=&T_R\bigg[\frac{2+N+N^2}{N(N+1)(N+2)}  \bigg]
        ~\to~\frac{T_R}{N},~{\rm and}\\
 P^{gg}(N)~=~\gamma_{gg}^{(1)}(N)
&=&2C_A\bigg[\frac{1}{N(N-1)}+\frac{1}{(N+1)(N+2)}-\sum_{k=1}^N
             \frac{1}{k}\bigg]+\beta_0
        ~\to~-2C_A\ln\bar{N}+\beta_0+\cO\lr\frac{1}{N}\rr
  \label{eq:pggn}
\eea
those which are not diagonal. Here, we have introduced the variable
$\bar{N}=Ne^{\gamma_E}$, where $\gamma_E$ is the Euler constant.

\subsection{Refactorization}

The resummation of the large logarithms in Eqs.\ (\ref{eq:paa}) and
(\ref{eq:kaa}) is based on the observation that the separation of the
non-perturbative PDFs $f_{a,b/A,B}$ and the perturbative partonic cross section
$\sigma_{ab}$ in Eq.\ (\ref{eq:HadFacN}) is not unambiguously defined. It is
in particular possible to re-factorize the partonic cross section
\cite{Sterman:1986aj}
\bea
  \sigma_{ab}(N,M^2,\mu^2)&=& H_{ab}(M^2,\mu^2)
  \frac{\psi_{a/a}(N,M^2)\psi_{b/b}(N,M^2)}{f_{a/a}(N,\mu^2)
  f_{b/b}(N,\mu^2)}S_{ab}(N,M^2) +\cO\lr\frac{1}{N}\rr
  \label{eq:TR:MasFac}
\eea
into a hard function $H_{ab}$, which is non-singular in and in fact independent of
$N$, a ratio of PDFs in partons $\psi_{a,b}$ and $f_{a,b}$ defined
at fixed {\em energy} and longitudinal momentum fraction, respectively, and a
function $S_{ab}$, which describes the emission of soft gluons and can thus be
computed in the eikonal approximation. The hard function
\bea
  H_{ab}(M^2,\mu^2)&=&\sum_{n=0}^\infty a_s^n(\mu^2) H_{ab}^{(n)}(M^2,\mu^2)
  \label{eqn:PR:HCoeff}
\eea
can be expanded as a power series in $a_s$. Its LO and NLO coefficients read
\bea
  H_{ab}^{(0)}(M^2,\mu^2)&=&\sigma_{ab}^{(0)}(M^2)~{\rm and}\\
  H_{ab}^{(1)}(M^2,\mu^2)&=&\sigma_{ab}^{(0)}(M^2)
  \le\cA_0+\lr\gamma_a^{(1)}+\gamma_b^{(1)}\rr
  \ln\frac{M^2}{\mu^2}\re,
\eea
where $\cA_0$ represents the IR-finite part of the renormalized virtual correction
\bea
  \sigma_{ab}^{(V)}(M^2,\mu^2) &=& a_s\bigg(\frac{4\pi\mu^2}{M^2}
  \bigg)^\eps \frac{\Gamma(1-\eps)}{\Gamma(1-2\eps)} \bigg(
  \frac{\cA_{-2}}{\eps^2}+\frac{\cA_{-1}}{\eps}+\cA_0\bigg) 
  \sigma_{ab}^{(0)}(M^2)+\cO(\eps).
  \label{eq:VirDef}
\eea
The PDFs satisfy the evolution equations
\cite{Altarelli:1977zs}
\bea
  \frac{\partial f_{c/a}(N,\mu^2)}{\partial\ln\mu^2}&=&
  \sum_b \gamma_{bc}(N,a_s(\mu^2))\,f_{b/a}(N,\mu^2),
  \label{eq:DGLAP}
\eea
governed by the (gauge-independent) anomalous dimensions $\gamma_{aa'}(N,a_s(\mu^2
))=\sum_n a_s^n(\mu^2)\gamma_{aa'}^{(n)}(N)$ of the composite Wilson operators
with $\gamma_{aa'}^{(1)}(N)$ given above, and \cite{Sterman:1986aj}
\bea
  \frac{\partial\psi_{a/a}(N,\mu^2)}{\partial\ln\mu^2}&=&
  \gamma_{a}(a_s(\mu^2)) \psi_{a/a}(N,\mu^2),
  \label{eq:DGLAP2}
\eea
where the (gauge-dependent) anomalous dimensions $\gamma_a=1/Z_a\,\partial Z_a/
\partial \ln\mu^2=\sum_n a_s^n(\mu^2)\gamma_{a}^{(n)}$ of the fields $a$
correspond
in the axial gauge \cite{Frazer:1978jd} to the $N$-independent (virtual) parts of
$\gamma_{aa}(N)$ \footnote{For a comparison with Eq.\ (\ref{eq:qwfc}) one must,
of course, evaluate $\delta Z_q$ in axial (not Feynman) gauge and recover the
renormalization scale dependence through the replacement $\ln4\pi \to\ln4\pi
\mu_R^2$.}. In the singlet/non-singlet basis and to one-loop order, the evolution
equations for $f_{c/a}$ can be solved and written in the closed exponential form
\cite{Furmanski:1981cw}
\bea
  f_{c/a}(N,\mu^2)&=&\sum_b E_{bc}(N,\mu^2,\mu_0^2) f_{b/a}(N,\mu_0^2),
  \label{eq:EvoSol}
\eea
where the evolution operator $E(N,\mu^2,\mu_0^2)$ satisfies the same evolution
equation, Eq.\ (\ref{eq:DGLAP}), as $f_{c/a}(N,\mu^2)$. The ratio
\bea
 {\psi_{a/a}(N,M^2)\over f_{a/a}(N,\mu^2)}&=&
 \Delta_a(N,M^2,\mu^2) \,U_a(N,M^2)^{1/2}
\eea
has been shown \cite{Sterman:1986aj} to exponentiate to all orders into a
gauge-independent, but scheme- and scale-dependent function
\bea
  \ln \Delta_a(N,M^2,\mu^2) &=&
  \int_0^1 \d z  \frac{z^{N-1}-1}{1-z} 
  \le \int_{(1-z)^{m_s}\mu^2}^{(1-z)^2M^2} \frac{\d q^2}{q^2} A_a(a_s(q^2))
  - B_a(a_s((1-z)^{m_s}M^2))\re
  \label{eq:TR:DeltaA}
\eea
with $m_s=0$ and consequently $B_a=0$ in the $\MSbar$-scheme, which describes the
soft and {\em collinear} gluon radiation from the initial partons, and a
gauge-dependent, but scheme- and scale independent function
\bea
 \ln U_a(N,M^2)&=&
 -\int_0^1 \d z  \frac{z^{N-1}-1}{1-z} 
 \nu_a(a_s((1-z)^2M^2)),
\eea
which can be combined with the soft function
\bea
 \ln S_{ab}(N,M^2)
 &=&+\int_0^1 \d z  \frac{z^{N-1}-1}{1-z} 
 \lambda_a(a_s((1-z)^2M^2))\delta_{ab}\!\!\!\!
\eea
into the gauge-, scheme- and scale-independent function
\bea
  \ln \Delta_{ab}(N,M^2) &=&
  \int_0^1 \d z  \frac{z^{N-1}-1}{1-z} 
  D_{ab}(a_s((1-z)^2M^2)).
  \label{eq:TR:DeltaD}
\eea
The functions $U_a$ and $S_{ab}$ can be computed in the eikonal approximation.
They depend on the cusp anomalous dimension $\nu_a=2C_a(\gamma\coth\gamma-1)$
\cite{Polyakov:1980ca} of the considered process and describe soft
{\em wide-angle} radiation.

\subsection{NLL approximation}

The coefficients
\bea
  A_a   (a_s)&=&\sum_{n=1}^\infty a_s^n A_a^{(n)},~~~
  B_a   (a_s)~=~\sum_{n=1}^\infty a_s^n B_a^{(n)},~~~{\rm and}~~~
  D_{ab}(a_s)~=~\sum_{n=1}^\infty a_s^n D_{ab}^{(n)}
  \label{eq:TR:ACoef}
\eea
can be expanded as power series in $a_s$, and the NLL results read
\cite{Sterman:1986aj,Catani:1989ne}
\bea
  A_a^{(1)}&=&2C_a,\\
  A_a^{(2)}&=&2C_a\bigg[\bigg(\frac{67}{18}-
  \frac{\pi^2}{6}\bigg)C_A -\frac{5}{9}N_f\bigg], \\
  B_a^{(1)}&=&0 ~({\rm in~the~}\MSbar~{\rm scheme}),~{\rm and} \\
  D_{ab}^{(1)}&=&0 ~({\rm since}~\nu_a^{(1)}=\lambda_a^{(1)})
  \label{eq:TR:NllCoeff}
\eea
with $\nu_a^{(1)}=4C_a$ and $C_a=C_F$ and $C_A$ for an incoming quark ($a=q$)
and gluon ($a=g$), respectively. Note, however, that $D_{ab}^{(2)}\neq 0$
\cite{Vogt:2000ci}.
After the integrations in Eqs.\ (\ref{eq:TR:DeltaA}) and (\ref{eq:TR:DeltaD}) have
been performed, the partonic cross section in \eqref{eq:TR:MasFac} can be written
in the closed exponential form \cite{Vogt:2000ci}
\bea
  \sigma_{ab}(N,M^2,\mu^2)&=& \cH_{ab}(M^2, \mu^2)
  \exp[\cG_{ab}(N,M^2,\mu^2)] +\cO\lr\frac{1}{N}\rr.
  \label{eq:TR:ExpTh2}
\eea
Here, the perturbative coefficients of the hard function
\bea
  \cH_{ab}(M^2,\mu^2)&=&\sum_{n=0}^\infty a_s^n(\mu^2) \cH_{ab}^{(n)}(M^2,\mu^2)
  \label{eq:TR:cHCoeff}
\eea
have been redefined with respect to those in \eqref{eqn:PR:HCoeff} in order to
absorb the non-logarithmic terms resulting from the integration, i.e.\
\bea
  \cH_{ab}^{(0)}(M^2, \mu^2)&=&\sigma^{(0)}_{ab}(M^2)~{\rm and}\\
  \cH_{ab}^{(1)}(M^2, \mu^2)&=&\sigma_{ab}^{(0)}(M^2)
  \le\cA_0
  +\frac{\pi^2}{6}\lr A^{(1)}_a+A^{(1)}_b\rr
  +\lr\gamma_a^{(1)}+\gamma_b^{(1)}\rr
  \ln\frac{M^2}{\mu^2}\re.
  \label{eq:TR:HcHRel}
\eea
The coefficient function $\cH_{ab}^{(1)}(M^2, \mu^2)$ given here agrees with the
one presented in Eq.\ (115) of Ref.\ \cite{Li:2007ih} except for their last three
terms. While their last term corresponds to the flavor-diagonal collinear
improvement to be discussed below, the two other terms represent leading and
next-to-leading logarithms and should therefore not be present. Furthermore, the
two terms in $(\ln 4\pi\mu_r^2/Q^2-\gamma_E)$ should be squared individually, not
together, and the virtual correction $\tilde{\cal M}_V^{\rm QCD}$ defined in
Eq.\ (116) of Ref.\ \cite{Li:2007ih} should include the complete SUSY-QCD contributions and not only their
UV-singular parts.
The function $\cG_{ab}$ takes the form
\bea
  \cG_{ab}(N,M^2,\mu^2)&=& L g_{ab}^{(1)}(\lambda) + 
  g_{ab}^{(2)}(\lambda,M^2/\mu^2) + a_s g_{ab}^{(3)}(\lambda,M^2/\mu^2) + \dots
  \label{eq:TR:cG}
\eea
with $\lambda = a_s \beta_0 L$ and $L = \ln \bar{N}$. The first term in
\eqref{eq:TR:cG},
\bea
 Lg_{ab}^{(1)}(\lambda)&=&{L\over 2\lambda\beta_0}
 (A_a^{(1)}+A_b^{(1)})
 \big[2\lambda+(1-2\lambda)\ln(1-2\lambda)\big],
\eea
collects the leading logarithmic (LL) large-$N$ contributions $L (a_s L)^n$
and depends on $A_{a,b}^{(1)}$ only. The coefficients $A_{a,b}^{(2)}$,
$A_{a,b}^{(1)}$ and $D_{ab}^{(1)}$ determine the function
\bea
  2\beta_0g_{ab}^{(2)}(\lambda,M^2/\mu^2)&=&(A_a^{(1)}+A_b^{(1)})
  \ln(1-2\lambda)\ln\frac{M^2}{\mu^2} \nonumber \\
  &+&(A_a^{(1)}+A_b^{(1)})\frac{\beta_1}{\beta_0^2}\big[2\lambda+\ln(1-2\lambda)
  +\frac{1}{2}\ln^2(1-2\lambda)\big] \nonumber \\
  &-&(A_a^{(2)}+A_b^{(2)})\frac{1}{\beta_0}\big[2\lambda+\ln(1-2\lambda)\big] 
  +D_{ab}^{(1)}\ln(1-2\lambda),
  \label{eq:TR:g1g2}
\eea
which resums the next-to-leading logarithmic (NLL) terms $(a_s L)^n$. Similarly,
the functions $g_{ab}^{(n+1)}$ resum the N$^n$LL terms and depend on the
coefficients $A_{a,b}^{(n+1)}$, $A_{a,b}^{(k)}$ and $D_{ab}^{(k)}$ with
$1 \leq k \leq n$.

\subsection{Collinear improvement}

Up to this point, we have systematically neglected all terms of $\cO(1/N)$.
However, since the dominant $1/N$-terms, i.e.\ those of the form $a_s^n L^{2n-1}/
N$, stem from the universal collinear radiation of initial state partons,
they are expected to exponentiate as well. This has been proven to
next-to-next-to-leading order for deep-inelastic scattering and Drell-Yan
type processes \cite{Kramer:1996iq} and can be achieved by making the replacement
(cf.\ Eq.\ (\ref{eq:TR:Tru}) below)
\begin{equation}
  \cH_{ab}^{(1)} \rightarrow \cH_{ab}^{(1)}+
  L {A_a^{(1)}+A_b^{(1)} \over N}{\cal H}_{ab}^{(0)},
  \label{eq:TR:KLSImp}
\end{equation}
i.e.\ by including the corresponding subleading terms of the {\em diagonal}
splitting functions $\gamma_{aa,bb}(N)$ in Eqs.\ (\ref{eq:pqqn}) and
(\ref{eq:pggn}). Carrying on with this argument, it is even possible to resum
the terms of ${\cal O}(1/N)$ coming from the diagonal {\em and} non-diagonal
splitting functions by identifying the terms \cite{Kulesza:2002rh}
\bea
 Lg_{ab}^{(1)}(\lambda)&=&{L\over 2\lambda\beta_0}
 (A_a^{(1)}+A_b^{(1)})
 \big[ (-2\lambda)         \ln(1-2\lambda)\big]+...
\eea
with the LL approximation of the QCD evolution operators
$E_{ab}$ 
defined in \eqref{eq:EvoSol} and then promoting the LL to
the full one-loop approximation $E^{(1)}_{ab}$.
The resummed cross section, \eqref{eq:TR:ExpTh2},
can then be written in a collinearly improved form as
\bea
 \sigma_{ab}(N,M^2,\mu^2)&=&\sum_{a',b'}
 E_{aa'}^{(1)}(N,M^2/\bar{N}^2,\mu^2_F)\,
 \tilde{\cal H}_{a'b'}(M^2,\mu^2_R)\,
 \exp[\tilde{\cal G}_{a'b'}(N,M^2,\mu^2_R)]\,
 E_{bb'}^{(1)}(N,M^2/\bar{N}^2,\mu^2_F),
 \label{eq:TR:gfd}
\eea
where the dependences on the factorization and renormalization scales have
been recovered explicitly, the collinearly improved hard coefficient function
$\tilde{\cal H}_{ab}$ is expanded as usual as a power series in $a_s(\mu^2)$ and
its LO and NLO coefficients read now
\bea
  \tilde{\cal H}_{ab}^{(0)}(M^2,\mu^2)&=&\sigma_{ab}^{(0)}(M^2)~{\rm and}\\
  \tilde{\cal H}_{ab}^{(1)}(M^2,\mu^2)&=&\sigma_{ab}^{(0)}(M^2)
  \le \cA_0+\frac{\pi^2}{6}\lr A_{a}^{(1)}+A_{b}^{(1)}\rr\re,
  \label{eq:TR:asdf}
\eea
and the Sudakov exponential function $\tilde{\cG}_{ab}$ is expanded in the same
way as $\cG_{ab}$ in Eq.\ (\ref{eq:TR:cG}) with
\bea
 L\tilde{g}_{ab}^{(1)}(\lambda)&=&{L\over 2\lambda\beta_0}
 (A_{a}^{(1)}+A_{b}^{(1)}) \big[2\lambda+\ln(1-2\lambda)\big] ~~{\rm and}\\
 2\beta_0 \tilde{g}_{ab}^{(2)}(\lambda,M^2/\mu^2)&=&(A_{a}^{(1)}+A_{b}^{(1)})
 \big[2\lambda+\ln(1-2\lambda)\big]\ln\frac{M^2}{\mu^2} \nonumber \\
 &+&(A_{a}^{(1)}+A_{b}^{(1)})\frac{\beta_1}{\beta_0^2}\big[2\lambda+\ln(1-2\lambda)
  +\frac{1}{2}\ln^2(1-2\lambda)\big] \nonumber \\
 &-&(A_{a}^{(2)}+A_{b}^{(2)})\frac{1}{\beta_0}\big[2\lambda+\ln(1-2\lambda)\big]
 \nonumber\\
 &+&(-2\gamma_{a}^{(1)}-2\gamma_{b}^{(1)}+D_{ab}^{(1)})\ln(1-2\lambda).
  \label{eq:TR:hatg1g2}
\eea
Here, the anomalous dimensions $\gamma_{a,b}^{(1)}$ in $\tilde{g}^{(2)}_{ab}(\lambda,
M^2/\mu^2)$ have been introduced to cancel the NLL terms in the one-loop
approximation $E_{ab}^{(1)}$ of the evolution operators.

For the Drell-Yan process, it has been suggested that also the constant terms
in the hard coefficient function $\tilde{\cal H}_{ab}^{(1)}(M^2,\mu^2)$ can be
exponentiated, since these terms factorize the complete Born cross section,
include finite remainders of the infrared singularities in the virtual
corrections and are thus related to the corresponding singularities in the
real corrections giving rise to the large logarithms \cite{Eynck:2003fn}.
While this choice is supported by an explicit two-loop
calculation \cite{Hamberg:1990np} and can be applied to other Drell-Yan like
processes \cite{Bozzi:2007qr}, gaugino pair production does not proceed through
a single $s$-channel diagram, and the virtual corrections thus factorize only at
the level of amplitudes, but not the full cross section. Resumming some or all of
the finite terms in the hard coefficient function $\tilde{\cal H}_{ab}^{(1)}(M^2,
\mu^2)$ into an exponential as in Ref.\ \cite{Li:2007ih} seems therefore not to be
justified in
this case.

\subsection{Matching and inverse Mellin transform}

As mentioned above, the large logarithms, which spoil the convergence of the
perturbative series and must be resummed to all orders, appear close to
production threshold. Conversely, the perturbative cross section should
be valid far from this threshold. To obtain a reliable prediction in
all kinematic regions, both results must be consistently matched through
\bea
  \sigma_{ab}&=&\sigma^{\rm(res.)}_{ab}+\sigma^{\rm(f.o.)}_{ab}-
  \sigma^{\rm(exp.)}_{ab},
  \label{eq:TR:Mat}
\eea
i.e.\ by subtracting from the sum of the resummed (res.) cross section in
\eqref{eq:TR:gfd} and the fixed order (f.o.) cross section in Eq.\ (\ref{eq:34})
their overlap. The latter can be obtained by expanding (exp.) the resummed cross
section to the same fixed order as the perturbative result. At $\cO(a_s)$, one
then obtains
\bea
  \sigma_{ab}^{\rm(exp.)}(N,M^2,\mu^2)&=&
  \tilde{\cal H}_{ab}^{(0)}(M^2,\mu^2)+a_s \tilde{\cal H}_{ab}^{(1)}(M^2,\mu^2)
  \nonumber \\
  &-&a_s\bigg(2L -\ln \frac{M^2}{\mu^2}\bigg) \sum_{c}
  \big[\gamma_{ac}^{(1)}(N) \tilde{\cal H}_{cb}^{(0)}(M^2,\mu^2)
       +\tilde{\cal H}_{ac}^{(0)}(M^2,\mu^2) \gamma_{bc}^{(1)}(N)
 \big]\nonumber \\
  &-&a_s \tilde{\cal H}_{ab}^{(0)}(M^2,\mu^2) \big[L^2(A_a^{(1)}+A_b^{(1)})
  -2L(\gamma_a^{(1)}+\gamma_b^{(1)})\big].
  \label{eq:TR:Tru}
\eea

After the resummed result and its perturbative expansion have been obtained
in Mellin $N$-space and multiplied with the $N$-moments of the PDFs according
to Eq.\ (\ref{eq:HadFacN}), an inverse Mellin transform
\bea
 M^2{\d\sigma_{AB}\over\d M^2}(\tau)&=&{1\over2\pi i}\int_{\cC_N}\d N 
 \tau^{-N} M^2{\d\sigma_{AB}(N)\over \d M^2}
\eea
must be performed in order to obtain the observed hadronic cross section as a
function of $\tau=M^2/S$. Special attention must be paid to the
singularities in the resummed exponents $\tilde{g}_{ab}^{(1,2)}$, which are situated at
$\lambda=1/2$ and are related to the Landau pole of the perturbative coupling
$a_s$. To avoid this pole as well as those in the Mellin moments of the PDFs
related to the small-$x$ (Regge) singularity $f_{a/A}(x,\mu_0^2)\propto x^\alpha
(1-x)^\beta$ with $\alpha<0$, we choose an integration contour $\cC_N$
according to the {\em principal value} procedure proposed in Ref.\
\cite{Contopanagos:1993yq} and the {\em minimal prescription} proposed in Ref.\
\cite{Catani:1996yz} and define two branches
\bea
  \cC_N:~~ N=C+ze^{\pm i\phi}~~{\rm with}~~ z\in[0,\infty[,
  \label{eq:IT:Nbra}
\eea
where the constant $C$ is chosen such that the singularities of the $N$-moments
of the PDFs lie to the left and the Landau pole to the right of the
integration contour. While formally the angle $\phi$ can be chosen in the range
$[\pi/2,\pi[$, the integral converges faster, if $\phi>\pi/2$.

\section{Numerical Results}
\label{sec:4}

We now turn to our numerical analysis of threshold resummation effects on the
production of various gaugino pairs at the Tevatron $p\bar{p}$-collider
($\sqrt{S}=1.96$ TeV) and the LHC $pp$-collider ($\sqrt{S}=7-14$ TeV). For the
masses and widths of the electroweak gauge bosons, we use the current values of
$m_Z=91.1876$ GeV and $m_W=80.403$ GeV. The squared sine of the electroweak
mixing angle
\bea
 \sin^2\theta_W &=& 1-{m_W^2\over m_Z^2}
\eea
and the electromagnetic fine structure constant
\bea
 \alpha &=& {\sqrt{2} G_F m_W^2 \sin^2\theta_W \over \pi}
\eea
can be calculated in the improved Born approximation using the world average
value of $G_F=1.16637\cdot 10^{-5}$ GeV$^{-2}$ for Fermi's coupling
constant \cite{Amsler:2008zzb}. The CKM-matrix is assumed to be diagonal, and
the top quark mass is taken to be 173.1 GeV \cite{:2009ec}. The strong coupling
constant is evaluated
in the one-loop and two-loop approximation for LO and NLO/NLL+NLO results,
respectively, with a value of $\Lambda_{\overline{\rm MS}}^{n_f=5}$
corresponding to the employed LO (CTEQ6.6L1) and NLO (CTEQ6.6M) parton densities
\cite{Nadolsky:2008zw}. For the resummed and expanded contributions, the
latter have been transformed numerically to Mellin $N$-space. When we
present spectra in the invariant mass $M$ of the gaugino pair, we identify
the unphysical scales $\mu_{F}=\mu_R=\mu$ with $M$, whereas for total
cross sections we identify them with the average mass of the two produced
gauginos. The remaining theoretical uncertainty is estimated by varying the
common scale $\mu$ about these central values by a factor of two up and
down and the parton densities 
along the 22 eigenvector directions defined by the CTEQ collaboration.

\subsection{Benchmark points}

The running electroweak couplings as well as the
physical masses of the SUSY particles and their mixing angles are computed
with the computer program SPheno 2.2.3 \cite{Porod:2003um}, which includes a
consistent calculation of the Higgs boson masses and all one-loop and
the dominant two-loop radiative corrections in the renormalization group
equations linking the restricted set of SUSY-breaking parameters at the gauge
coupling unification scale to the complete set of observable SUSY masses and
mixing
angles at the electroweak scale. We choose the widely used minimal supergravity
(mSUGRA) point SPS1a' \cite{AguilarSaavedra:2005pw} as the benchmark for most
of our numerical studies. This point has an intermediate value of $\tan
\beta=10$ and $\mu>0$ (favored by the rare decay $b\to s\gamma$ and the measured
anomalous magnetic moment of the muon), a light gaugino mass parameter of $m_{1/2}
=250$ GeV, and a slightly lower scalar mass parameter $m_0=70$ GeV and trilinear
coupling $A_0=-300$ GeV than the original point SPS1a \cite{Allanach:2002nj} in
order to render it compatible with low-energy precision data, high-energy mass
bounds, and the observed cold dark matter relic density. It is also similar to
the post-WMAP point B' ($m_0=60$ GeV and $A_0=0$) \cite{Battaglia:2003ab}, which
has been adopted by the CMS collaboration as their first low-mass point (LM1)
\cite{Ball:2007zza}. In the SPS1a' scenario, the $\na$ is the LSP with a mass of
98 GeV, the gauginos producing the trilepton signal have masses of $m_{\ca}\simeq
m_{\nb}=184$ GeV, and the heavier gauginos, which decay mostly into the lighter
gauginos, $W$ and $Z$ bosons as well as the lightest Higgs boson, have masses of
$m_{\nc}=400$ GeV and
$m_{\cb}\simeq m_{\nd}=415$ GeV. The average squark and gluino masses are
$m_{\tilde{q}}\simeq 550$ GeV and $\mgl=604$ GeV.

Apart from the low-mass point LM1, we will also study the points LM7 and LM9,
since all three points have been found by the CMS collaboration to lead to visible
three-lepton signals. For LM7, the direct $\ca\nb$ production cross section
exceeds even 70\% of the total SUSY particle production cross section
\cite{Ball:2007zza}. The ATLAS collaboration have studied the
direct production of gauginos at the points SU2 and SU3 with or without a jet
veto (denoted JV, i.e.\ no jet in the event with transverse momentum $p_T>20$
GeV) in order to suppress the background from top quark pair production
\cite{Aad:2009wy}. A summary of all scenarios considered here is presented in
Tab.\ \ref{tab:2}. Note that none of these points falls into (but most of
\begin{table}[t]
 \centering
 \caption{Names, mSUGRA parameters, and physical SUSY particle masses of the
 benchmark points used in our numerical studies.\\}
 \begin{tabular}{|c|ccccc|cccccc|}
 \hline
 Scenario& $m_0$\,[GeV]\!& $m_{1/2}$\,[GeV]\!& $A_0$\,[GeV]\!& $\tan\beta$& sgn$(\mu)$ &
 $m_{\na}$\,[GeV]\! & $m_{\ca,\nb}$\,[GeV]\! & $m_{\nc}$\,[GeV]\! & $m_{\cb,\nd}$\,[GeV]\! & $m_{\tilde{q}}$\,[GeV]\! & $m_{\tilde{g}}$\,[GeV]\! \\
 \hline \hline
 SPS 1a'  & ~~~70     & 250            & $-$300     & 10         & $+$         &
 ~~98     & 184      & 400      & 415      & 550      & 604      \\
 \hline
 LM1      & ~~~60     & 250            &~~~~~0      & 10         & $+$         &
 ~~96     & 178      & 346      & 366      & 550      & 603      \\
 LM7      & 3000      & 230            &~~~~~0      & 10         & $+$         &
 ~~94     & 176      & 337      & 359      & 3000     & 636      \\
 LM9      & 1450      & 175            &~~~~~0      & 50         & $+$         &
 ~~70     & 128      & 263      & 284      & 1480     & 487      \\
 \hline
 SU2      & 3550      & 300            &~~~~~0      & 10         & $+$         &
 124      & 229      & 355      & 384      & 3560     & 809      \\
 SU3      & ~~100     & 300            & $-$300     & ~~6        & $+$         &
 118      & 223      & 465      & 481      & 650      & 715      \\
 \hline
 \end{tabular}
 \label{tab:2}
\end{table}
them lie relatively close to) the regions excluded by the Tevatron collaborations
CDF and D0, which assume, however, a lower value of $\tan\beta=3$ and always
$A_0=0$ \cite{Aaltonen:2008pv}. In Ref.\ \cite{Li:2007ih}, the cross section for
the associated production of $\ca$ and $\nb$ has been computed as a function
of $\tan\beta$ and $m_{1/2}$ for $m_0=200$ and 1000 GeV and assuming $A_0=0$ and
$\mu>0$. Unfortunately, the exact version of the renormalization group program
SPheno used there could not be determined, and we were not
able to reproduce the physical SUSY particle mass spectra of Ref.\
\cite{Li:2007ih}. Since we also do not completely agree analytically with the
coefficient function $\cH_{ab}^{(1)}(M^2, \mu^2)$ of Ref.\ \cite{Li:2007ih} (see
above), we must refrain from a direct comparison of our numerical results.

\subsection{Invariant mass spectra}

\begin{figure}[t]
  \includegraphics[width=.49\textwidth]{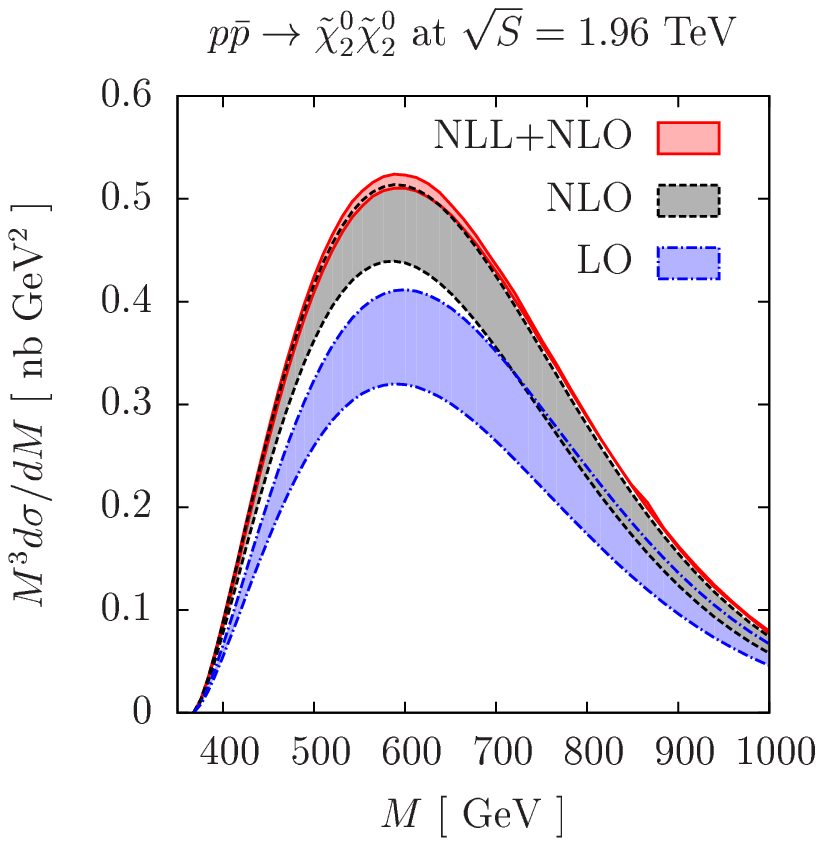}
  \includegraphics[width=.49\textwidth]{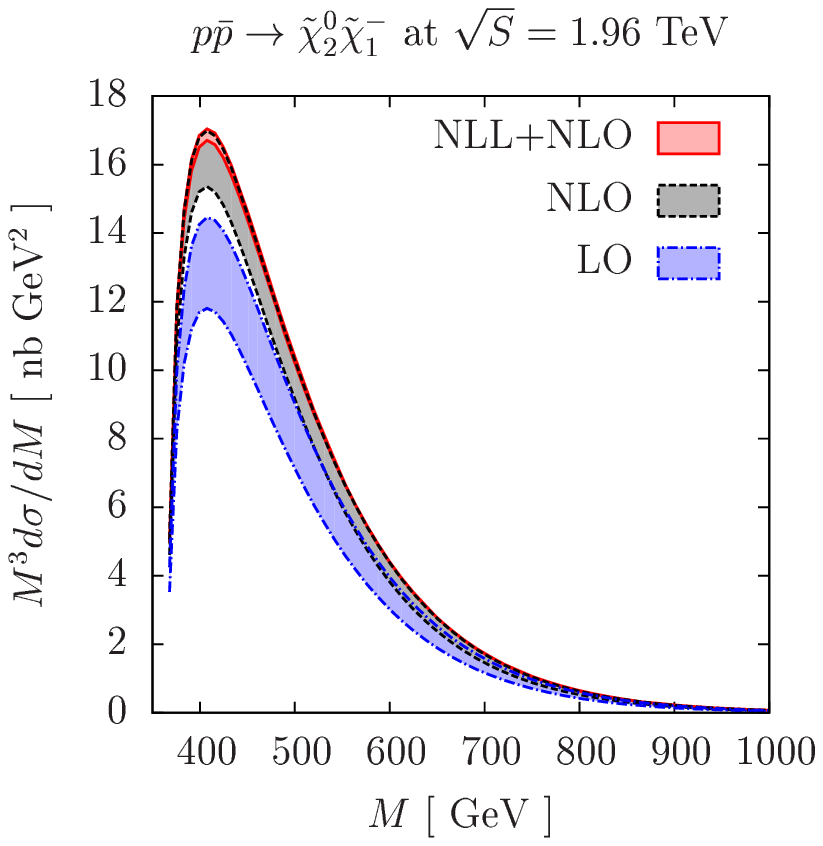}
  \includegraphics[width=.49\textwidth]{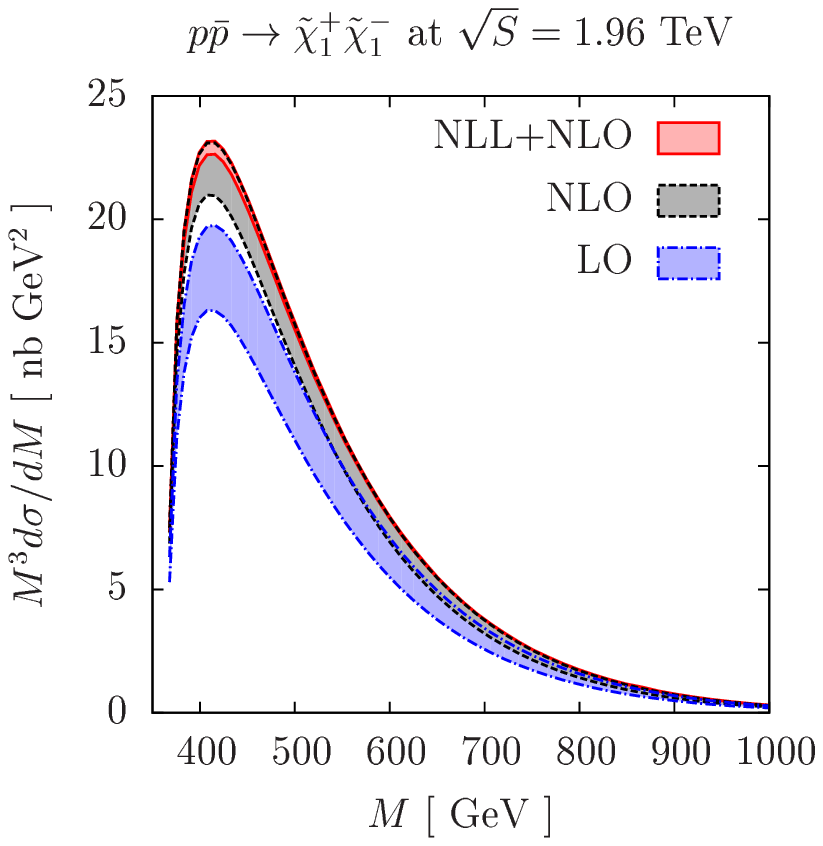}
  \caption{\label{fig:9}Invariant mass spectra for the production of various
 light gaugino pairs at the Tevatron in the SPS1a' scenario and in the LO (blue),
 NLO (grey) and NLL+NLO (red) approximation. The corresponding scale
 uncertainties are represented by the band widths.}
\end{figure}

In Fig.\ \ref{fig:9} we present invariant mass spectra $M^3\d\sigma/\d M$ for
the production of various combinations of $\ca$ and $\nb$ with $m_{\ca}\simeq
m_{\nb}=184$ GeV in the SPS1a' scenario at the Tevatron. The spectra start at
$M=m_{\ca}+m_{\nb}=368$ GeV and increase considerably from LO (blue) to NLO
(grey), but much less from NLO to NLL+NLO (red). The scale uncertainty is
considerably reduced from NLO to NLL+NLO, which indicates good convergence of
the reorganized perturbative series. The cross section is smallest for the
production of two neutralinos, since they are gaugino-like and couple only
weakly to the $s$-channel $Z$-boson (see Tab.\ \ref{tab:1}). Since the Tevatron
is a $p\bar{p}$ collider, the cross sections are identical for $\nb
\tilde{\chi}^-_1$ and $\tilde{\chi}^+_1\nb$ (not shown) pairs. The largest cross
section is obtained for
chargino pairs due to the $s$-channel photon contribution. Threshold resummation
should be most important as $M\to \sqrt{s}$ and $z\to1$, but its effects on the
partonic cross section are, of course, reduced in the hadronic cross section shown
here by the parton densities, which tend to $0$ as $x_{a,b}$ and $z\to1$.
Nevertheless, on close inspection one observes that the NLL+NLO cross section for
two neutralinos no longer overlaps with the one at NLO for relatively large
invariant masses of $M\simeq \sqrt{S}/2$.

\begin{figure}[t]
  \includegraphics[width=.49\textwidth]{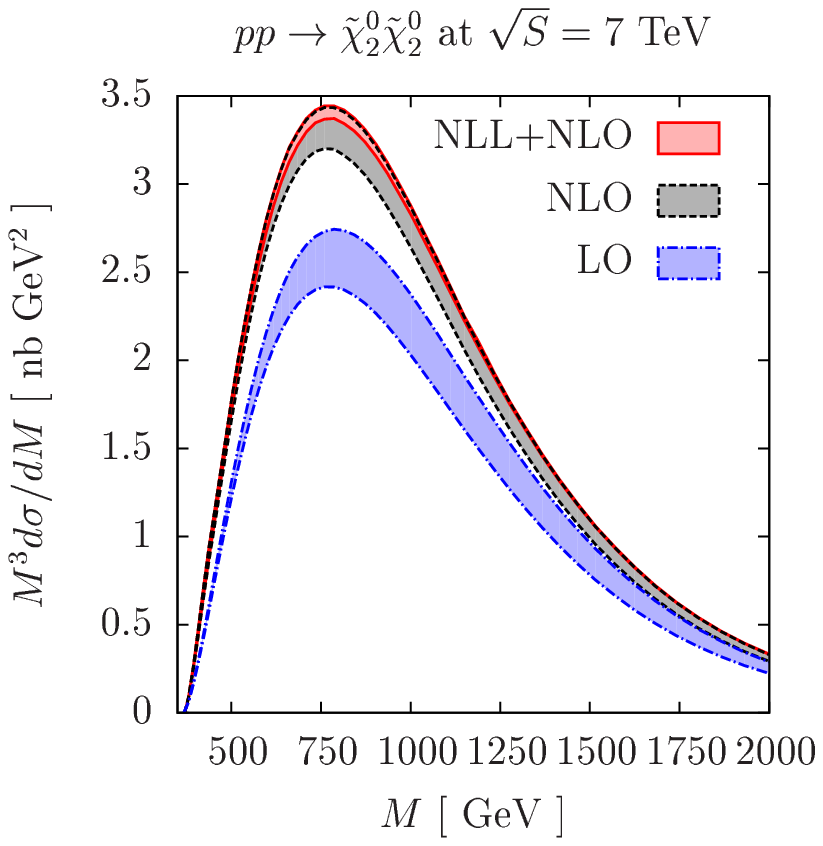}
  \includegraphics[width=.49\textwidth]{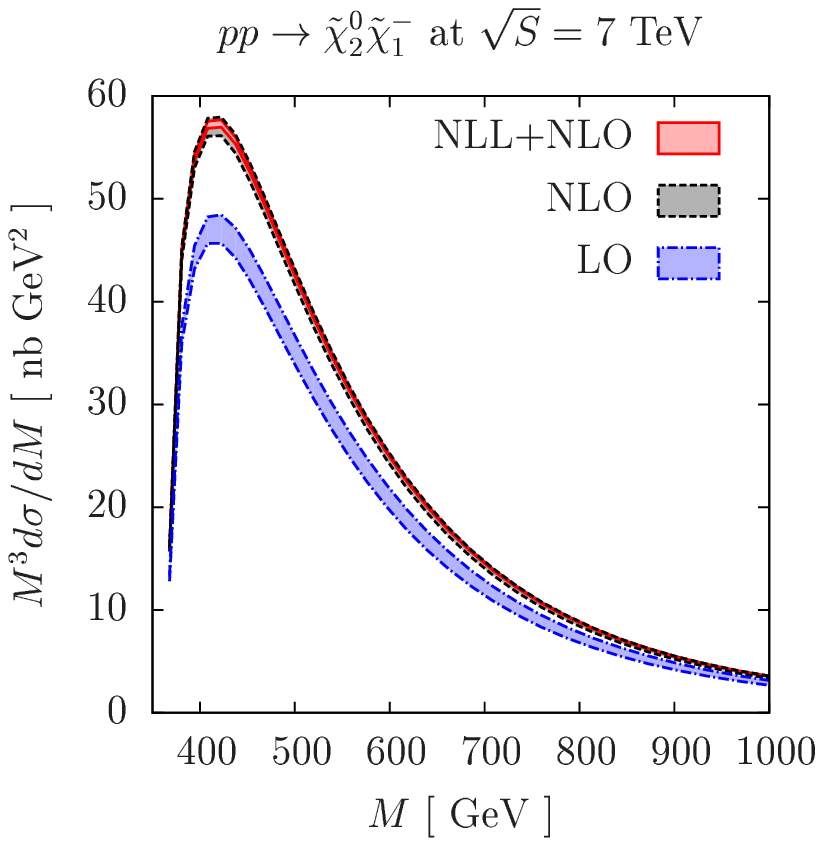}
  \includegraphics[width=.49\textwidth]{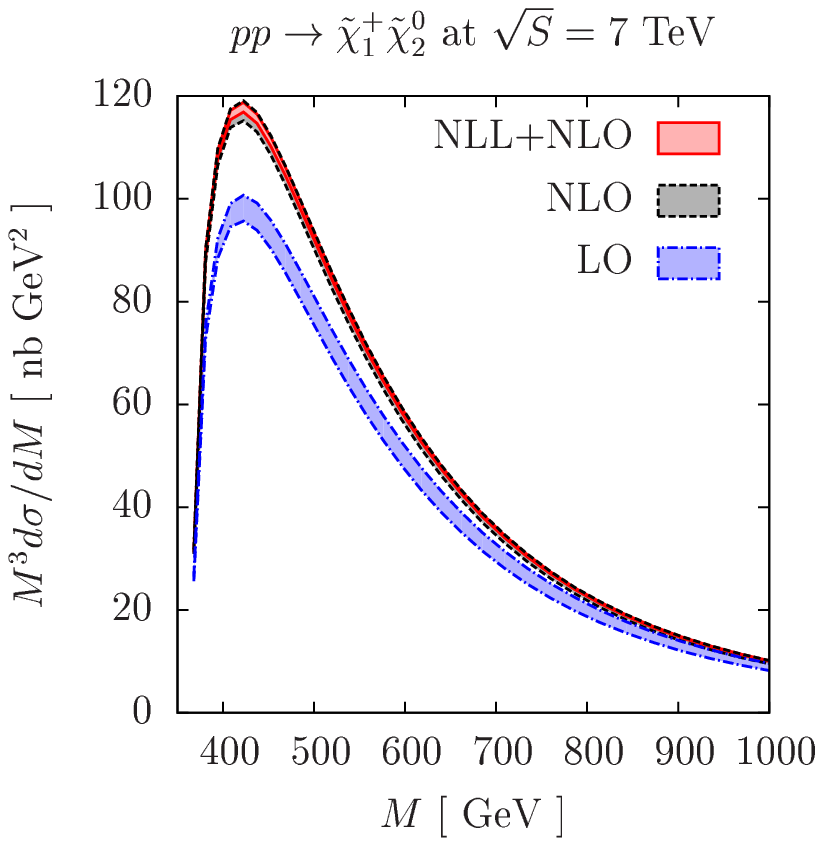}
  \includegraphics[width=.49\textwidth]{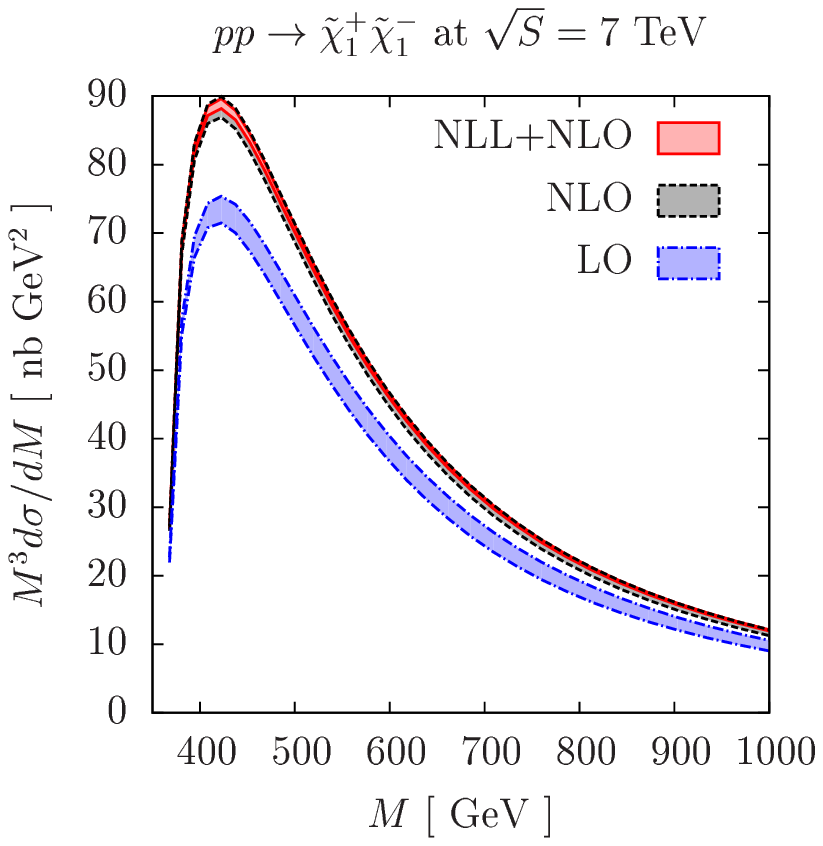}
  \caption{\label{fig:10}Same as Fig.\ \ref{fig:9} for the LHC with its current
 center-of-mass energy of $\sqrt{S}=7$ TeV.}
\end{figure}

\begin{figure}[t]
  \includegraphics[width=.49\textwidth]{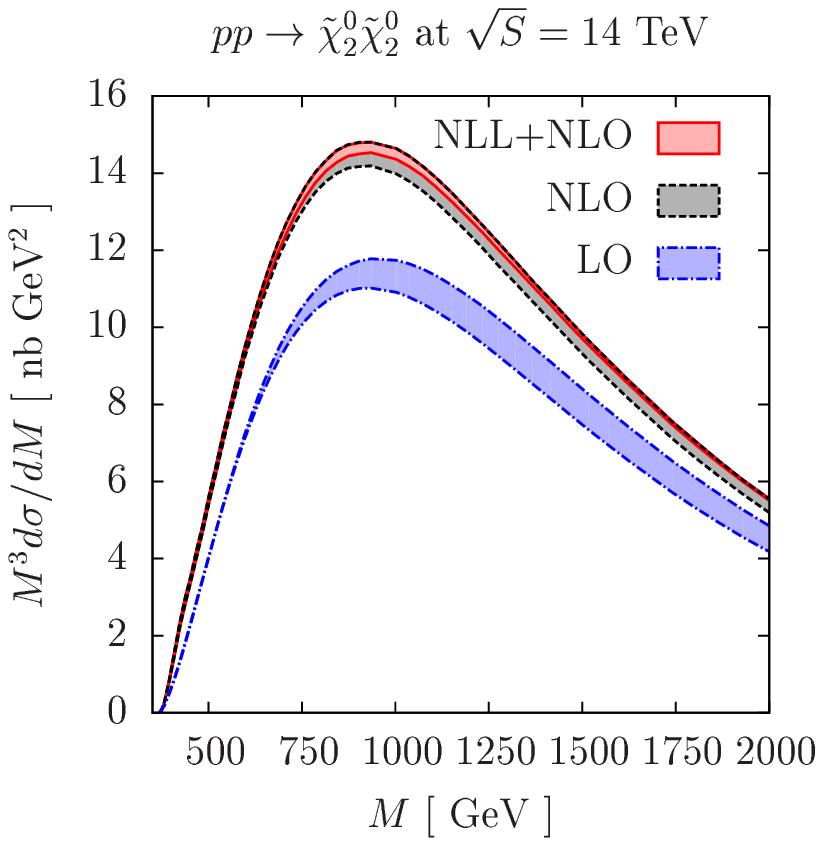}
  \includegraphics[width=.49\textwidth]{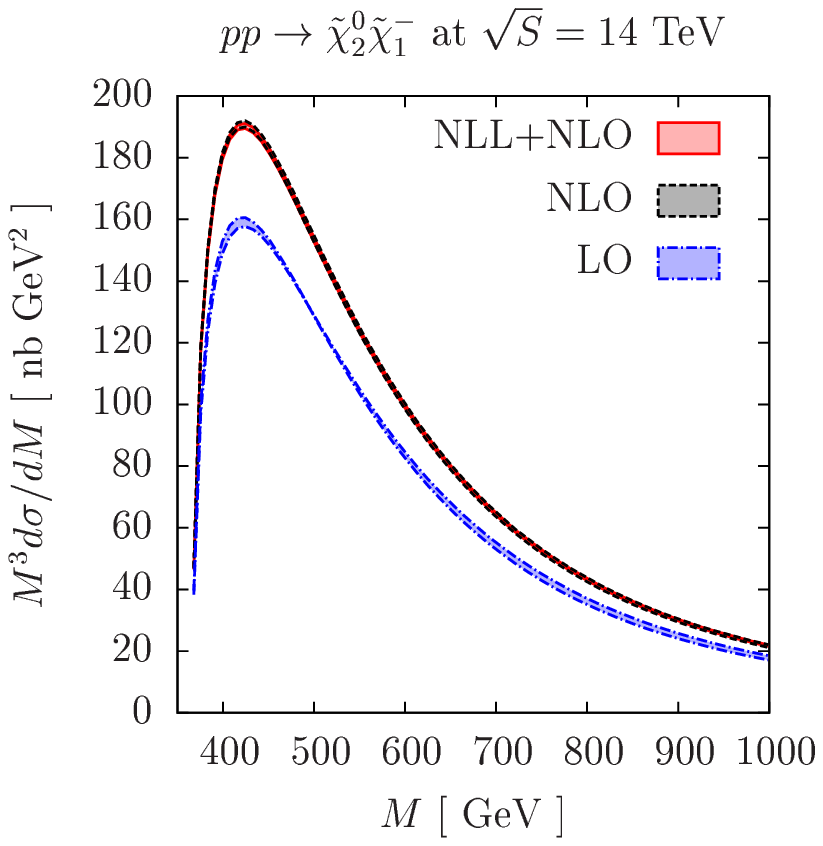}
  \includegraphics[width=.49\textwidth]{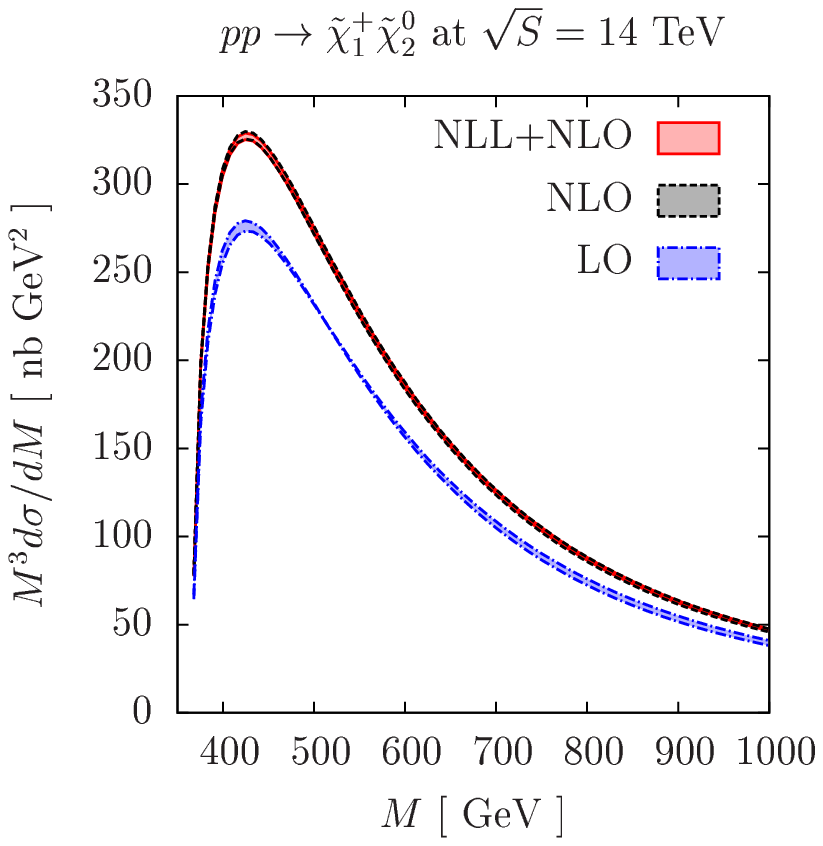}
  \includegraphics[width=.49\textwidth]{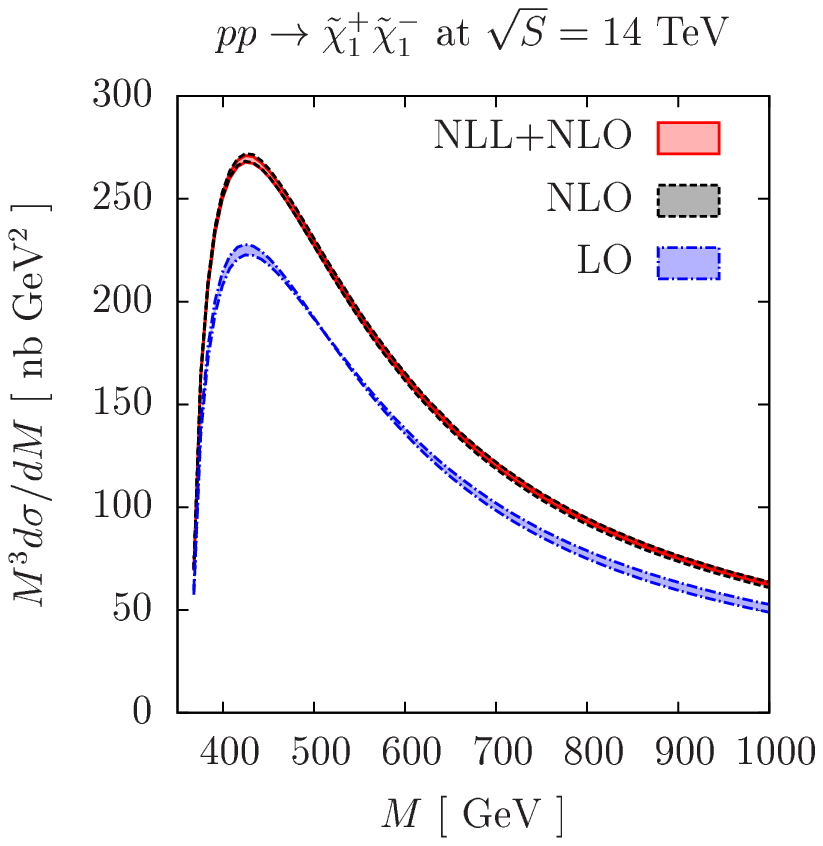}
  \caption{\label{fig:11}Same as Fig.\ \ref{fig:9} for the LHC with its design
 center-of-mass energy of $\sqrt{S}=14$ TeV.}
\end{figure}

A similar hierarchy of the different production channels is observed in Fig.\
\ref{fig:10} for the LHC with its
current center-of-mass energy of $\sqrt{S}=7$ TeV and in Fig.\ \ref{fig:11} for
the LHC with its design energy of $\sqrt{S}=14$ TeV. There are, however, two
notable differences. First, the LHC is a $pp$ collider, so that the cross section
for $\tilde{\chi}^+_1\nb$ exceeds the one for $\tilde{\chi}^-_1\nb$ by a factor
of two and becomes even larger than the one for chargino pairs. Second, the
NLO band is separated by a wider gap from the LO band than it was the case at
Tevatron, whereas the NLL+NLO and NLO bands overlap considerably more. This is,
of course, due to the fact that the light gauginos are now produced further away
from the threshold of the 7 or 14 TeV collider, so that the importance of soft-gluon
resummation is reduced. However, one still observes a sizeable reduction of the
scale uncertainty from NLO to NLL+NLO.

\begin{figure}[t]
  \includegraphics[width=.49\textwidth]{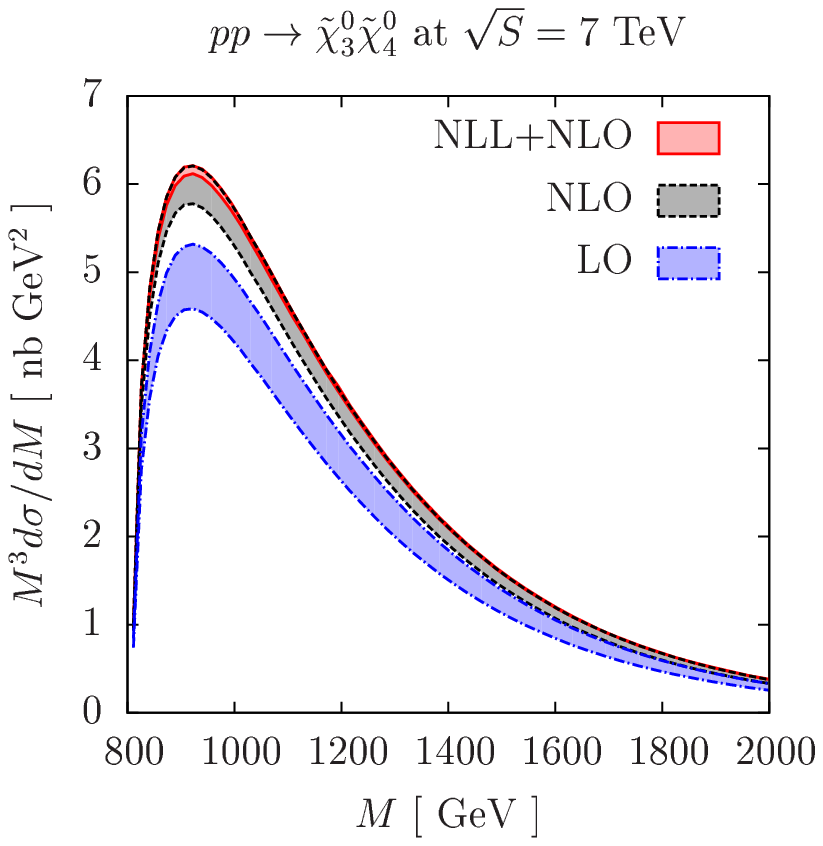}
  \includegraphics[width=.49\textwidth]{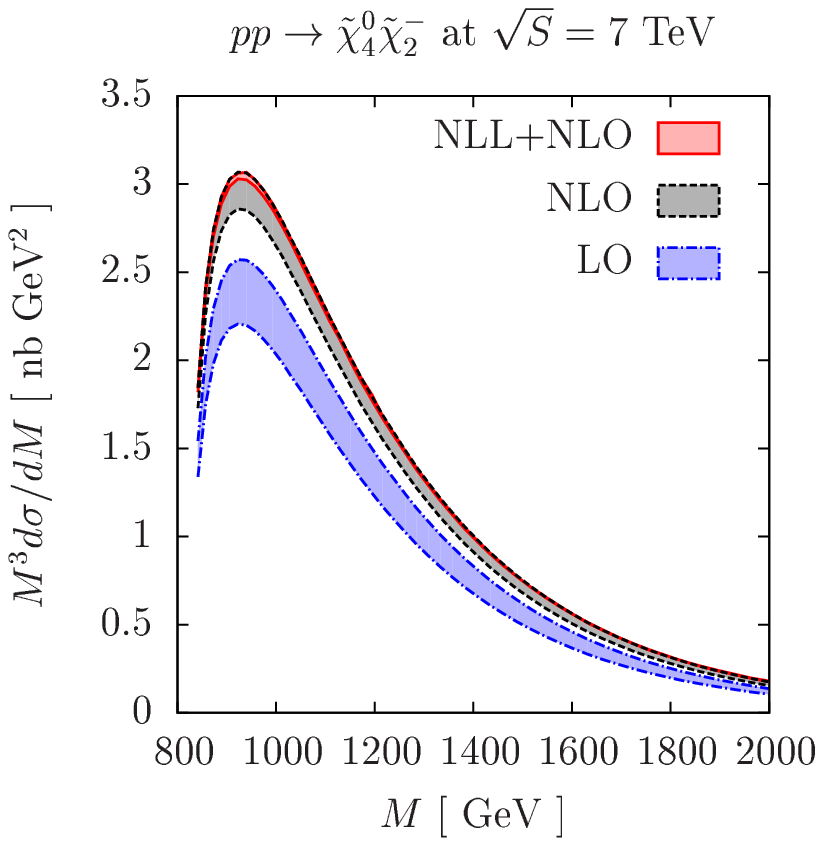}
  \includegraphics[width=.49\textwidth]{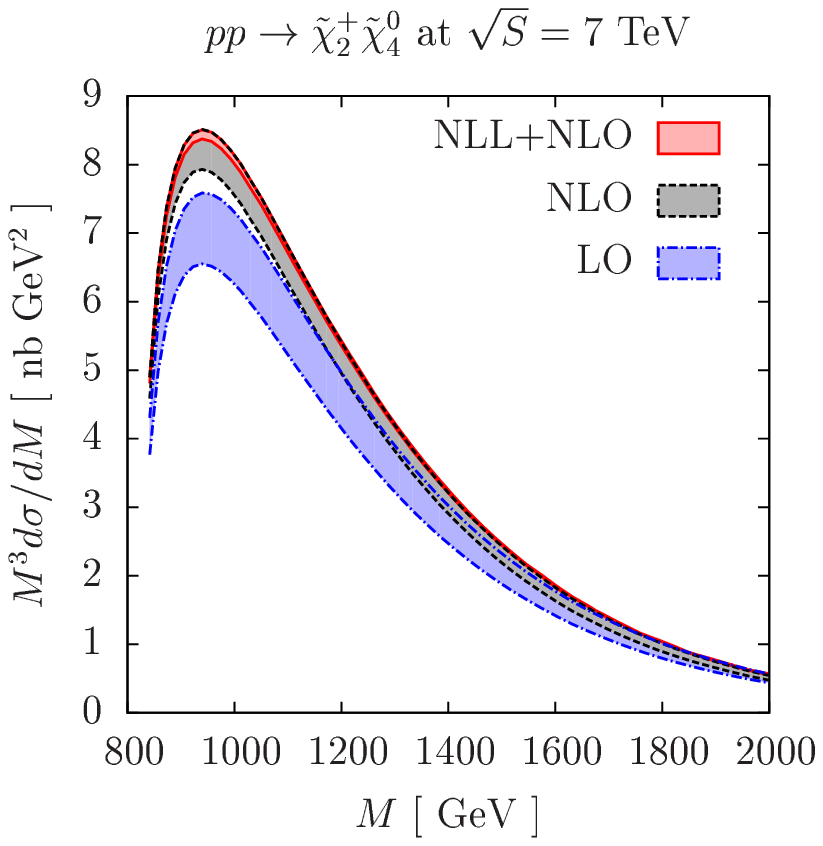}
  \includegraphics[width=.49\textwidth]{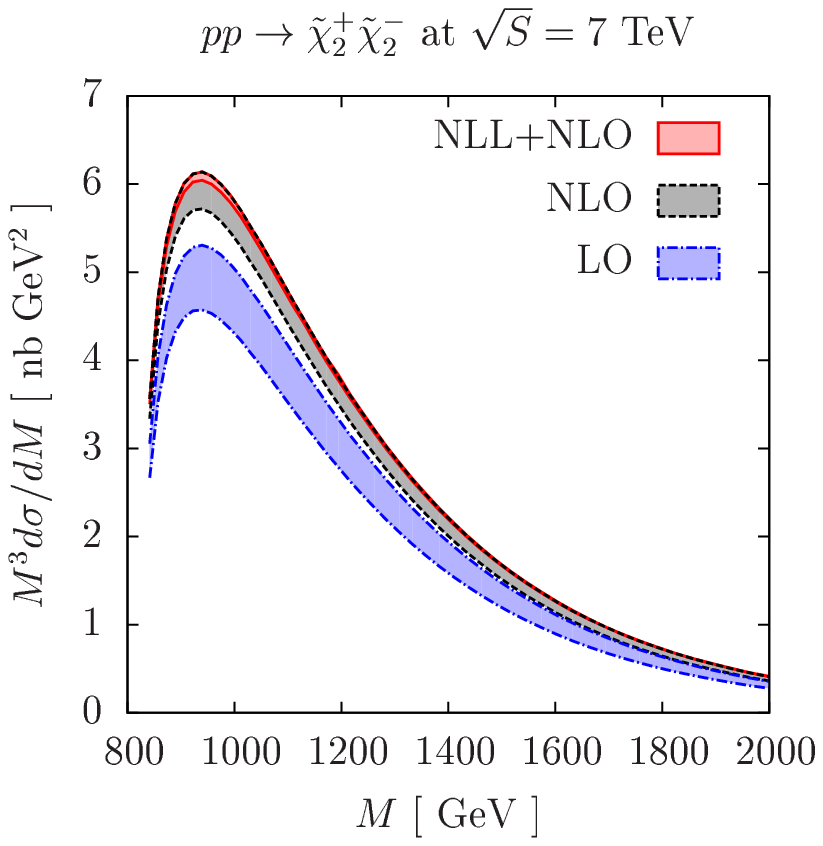}
  \caption{\label{fig:12}Same as Fig.\ \ref{fig:9} for the production of heavy
 gaugino pairs at the LHC with its current center-of-mass energy of $\sqrt{S}=7$
 TeV.}
\end{figure}

\begin{figure}[t]
  \includegraphics[width=.49\textwidth]{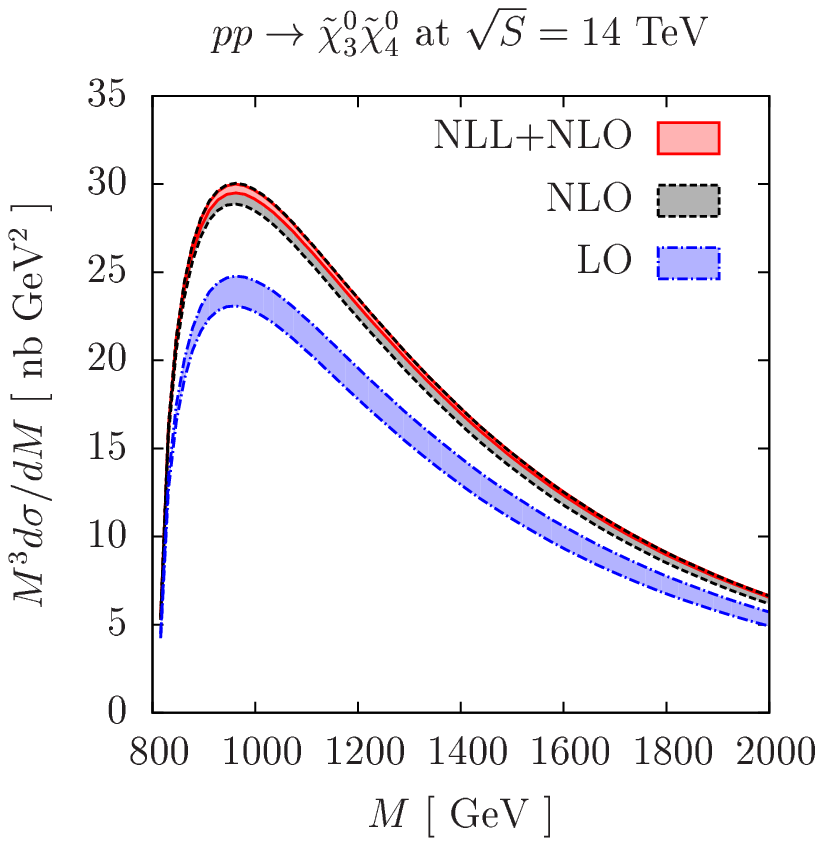}
  \includegraphics[width=.49\textwidth]{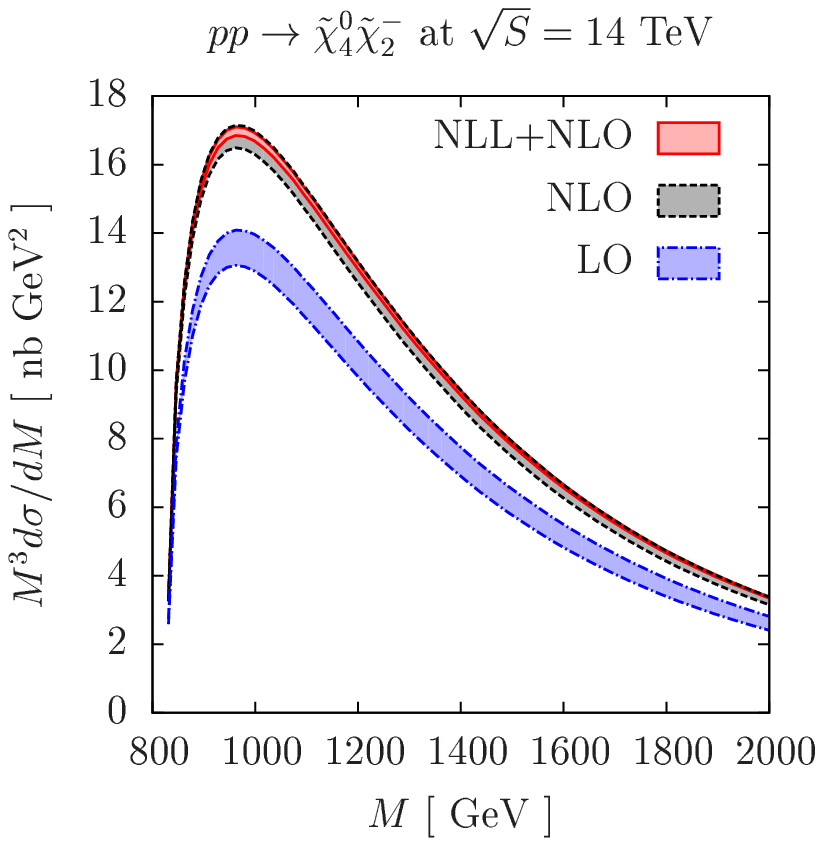}
  \includegraphics[width=.49\textwidth]{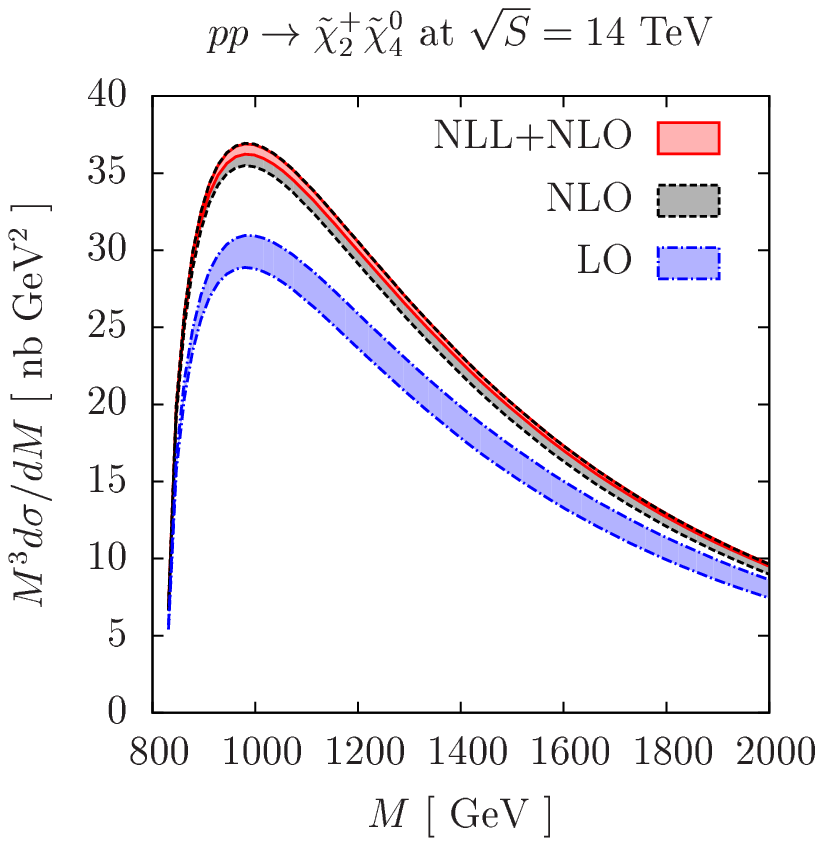}
  \includegraphics[width=.49\textwidth]{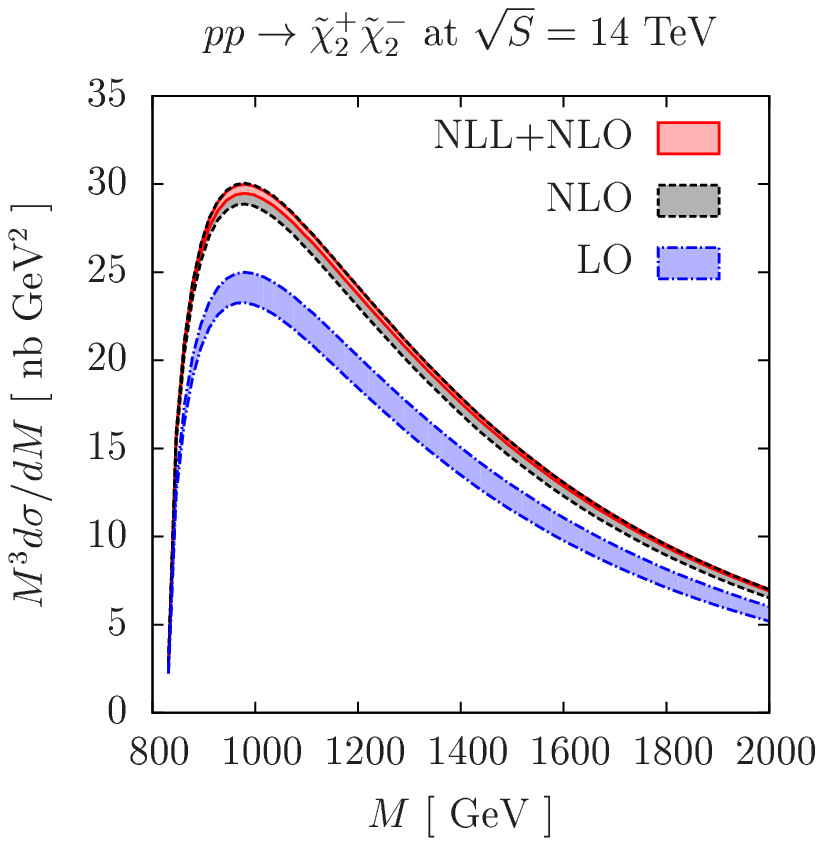}
  \caption{\label{fig:13}Same as Fig.\ \ref{fig:9} for the production of heavy
 gaugino pairs at the LHC with its design center-of-mass energy of $\sqrt{S}=14$
 TeV.}
\end{figure}

Heavier gaugino pairs can only be produced with sizeable cross sections at the
LHC. We therefore show in Figs.\ \ref{fig:12} and \ref{fig:13} the invariant mass
spectra $M^3\d\sigma/\d M$ for the production of various combinations of
$\tilde{\chi}^0_{3,4}$ and $\cb$ at the LHC with $\sqrt{S}=7$ TeV and 14 TeV
and with $m_{\nc}=400$ GeV and
$m_{\cb}\simeq m_{\nd}=415$ GeV in the SPS1a' scenario. The spectra start at $M
\simeq 800-830$ GeV, and their magnitudes are considerably smaller than in the
light gaugino case.
However, they are now of comparable size for neutralino and chargino pairs due
to the fact that the dominantly higgsino $\nc$ and $\nd$ now have sizeable
couplings to the $s$-channel $Z$-boson (see Tab.\ \ref{tab:1}). The associated
production of a neutralino and a chargino is again much larger for the positive
chargino eigenstate than for its negative counterpart. The cross sections for
$\nc\cb$ pairs are very similar to those for $\nd\cb$ pairs and therefore not
shown. Higgsino-like neutralinos and charginos with large $s$-channel
contributions are produced as $S$-waves, so that the invariant mass spectra rise
more steeply at low $M$ than $P$-wave produced gaugino-like neutralinos and
charginos.

\begin{figure}[t]
  \includegraphics[width=.49\textwidth]{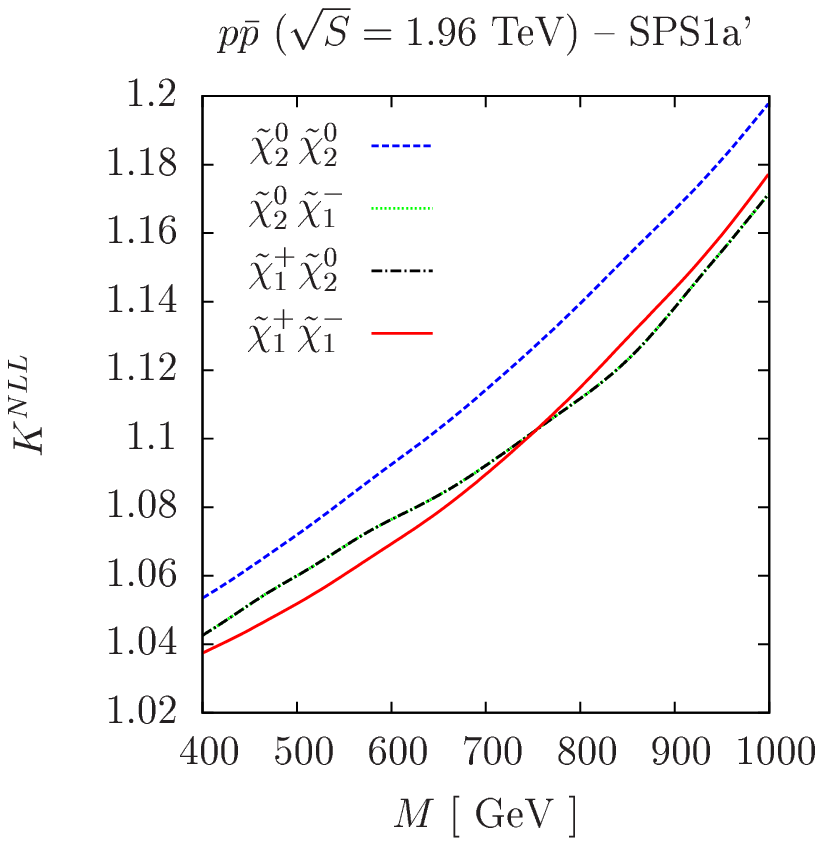}
  \includegraphics[width=.49\textwidth]{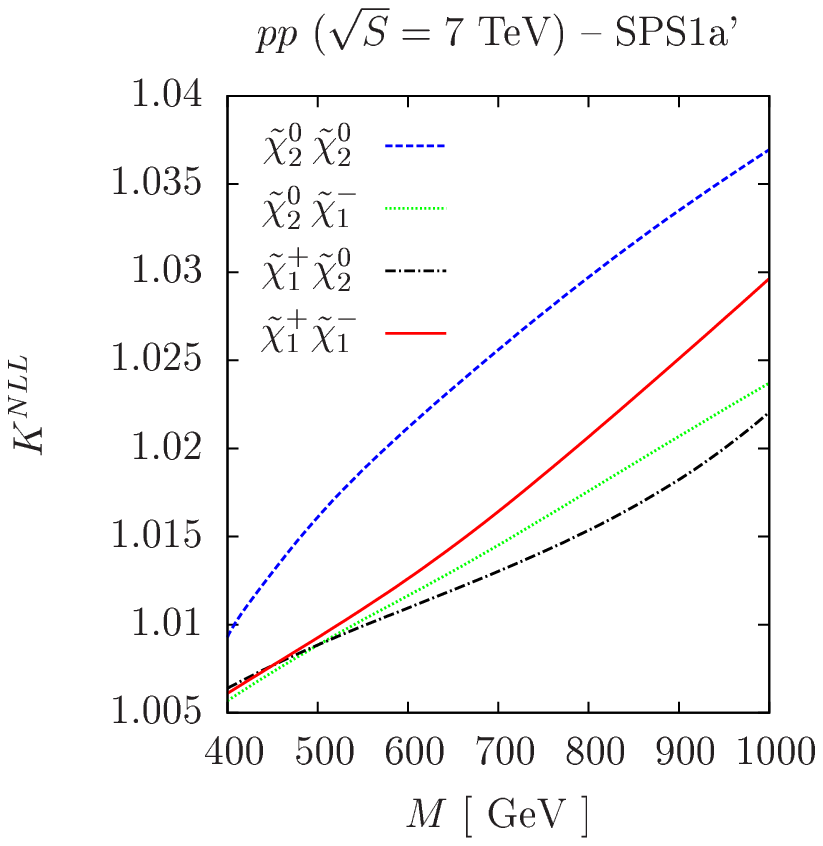}
  \includegraphics[width=.49\textwidth]{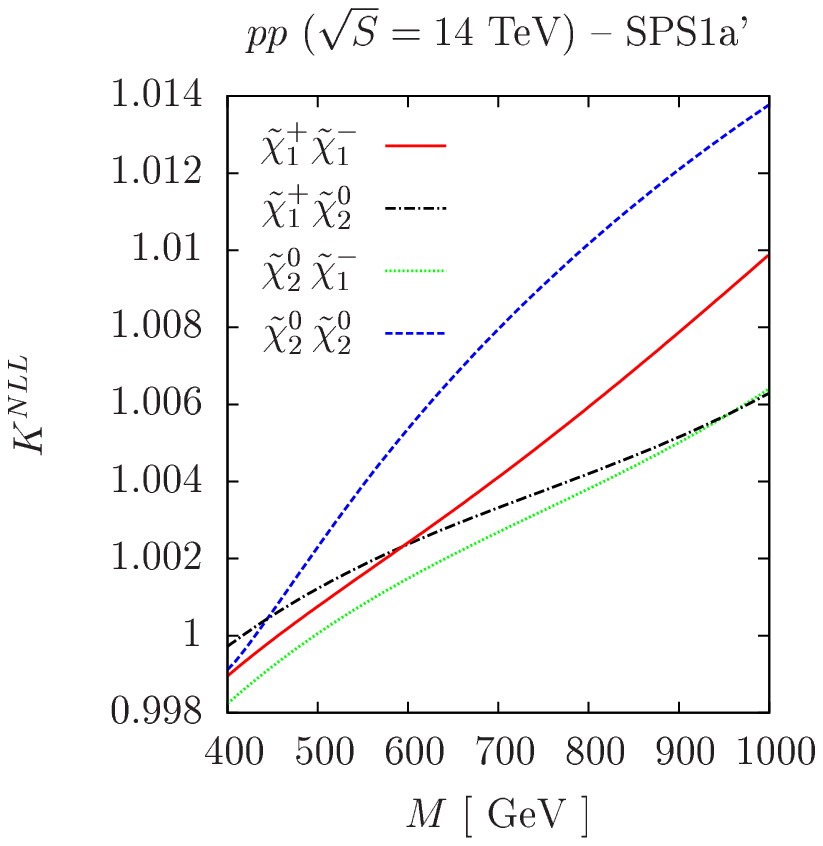}
  \caption{\label{fig:14}Ratios $K^{\rm NLL}$ of NLL+NLO over NLO differential
 cross sections as a function of the invariant mass $M$ of a light gaugino pair
 at the Tevatron (top left) and LHC with $\sqrt{S}=7$ TeV (top right) and
 $\sqrt{S}=14$ TeV (bottom) in the SPS1a' scenario.}
\end{figure}

From Figs.\ \ref{fig:9}-\ref{fig:13}, the impact of threshold resummation effects
is difficult to estimate. We therefore present in Fig.\ \ref{fig:14} the relative
size
\begin{equation}
  K^{\rm NLL} = \frac{d \sigma^{\rm NLL+NLO}}{d \sigma^{\rm NLO}}
\end{equation}
of the NLL+NLO prediction with respect to the NLO prediction. As one expects,
the correction is larger at the Tevatron with its lower center-of-mass energy
(top left) than at the LHC (top right and bottom) and increases with the
invariant mass. The relatively small differences among the $K^{\rm NLL}$-factors
for neutralino pair production and the channels involving at least one chargino
can be traced to the fact that the former receives most of its contributions
from $t$- and $u$-channel squark exchanges, which are more sensible to strong
corrections than the exchanges of electroweak bosons in the $s$-channel.

\begin{figure}[t]
  \includegraphics[width=.49\textwidth]{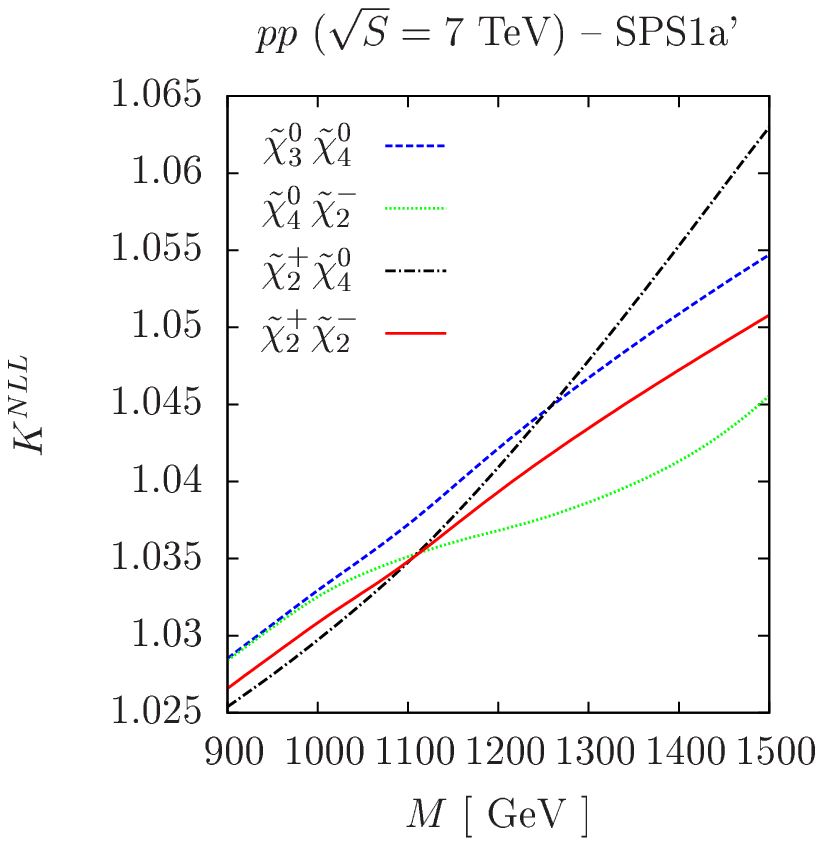}
  \includegraphics[width=.49\textwidth]{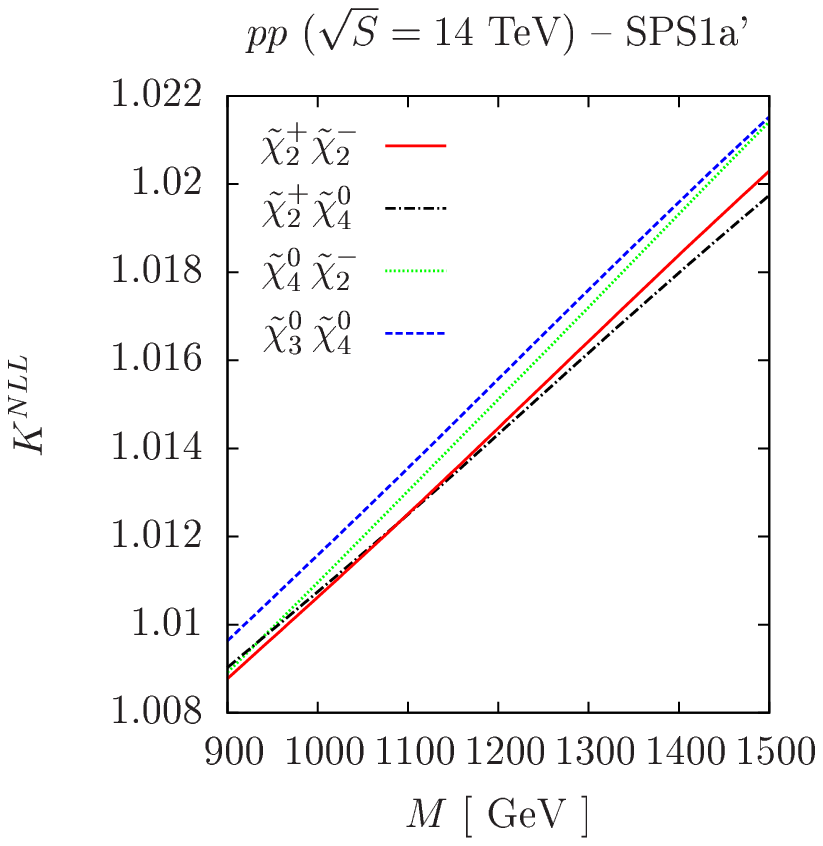}
  \caption{\label{fig:15}Same as Fig.\ \ref{fig:14} for heavy gaugino pairs
 at the LHC with $\sqrt{S}=7$ TeV (left) and $\sqrt{S}=14$ TeV (right).}
\end{figure}

The $K^{\rm NLL}$-factors for the production of heavy gaugino pairs at the LHC
with $\sqrt{S}=7$ TeV (14 TeV) are presented in Fig.\ \ref{fig:15} left (right).
They are larger than their counterparts for light gauginos in Fig.\ \ref{fig:14},
since the gaugino masses as well as the invariant masses $M$ are now closer to the
hadronic center-of-mass energies. In addition, the result for the $\nc\nd$ channel
differs no longer substantially from the other channels, since the heavy
neutralinos are now higgsino-like and their production is now also dominated by
the $s$-channel exchange of a weak gauge boson.

\subsection{Total cross sections}

\begin{figure}[t]
  \centering
  \includegraphics[width=.49\textwidth]{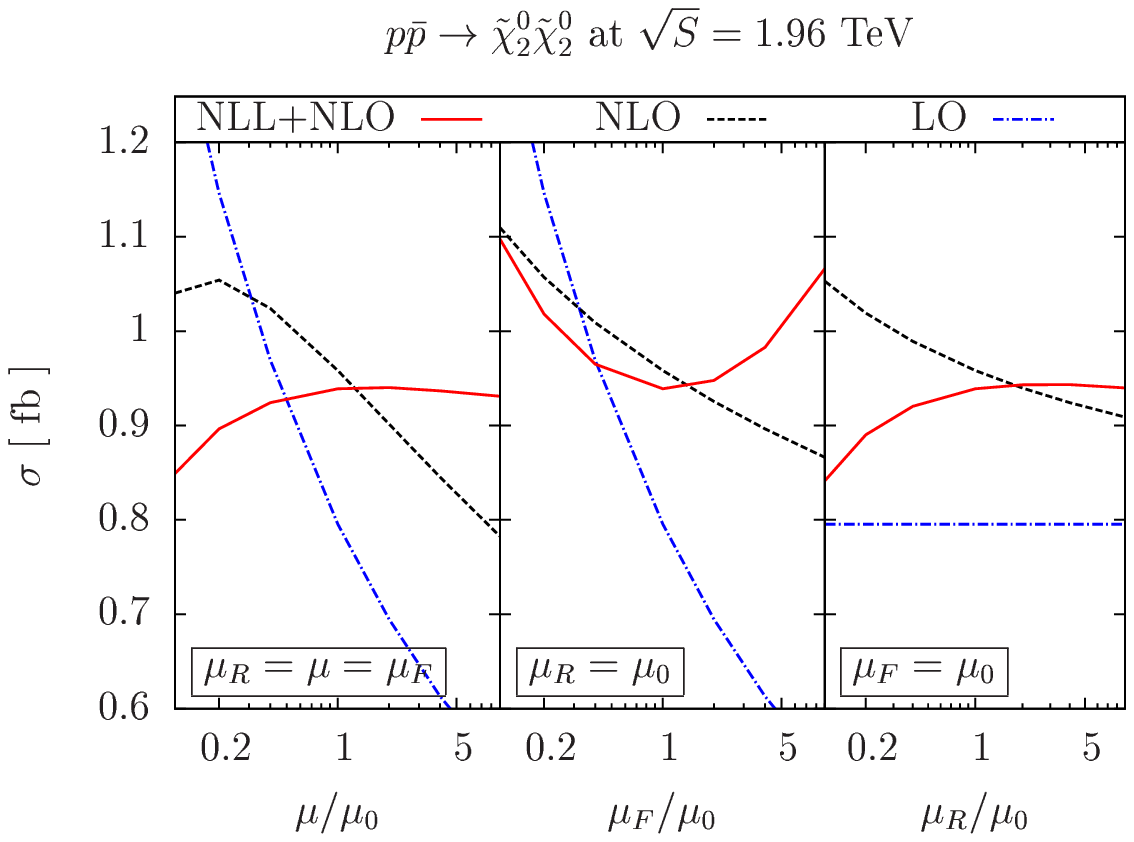}
  \includegraphics[width=.49\textwidth]{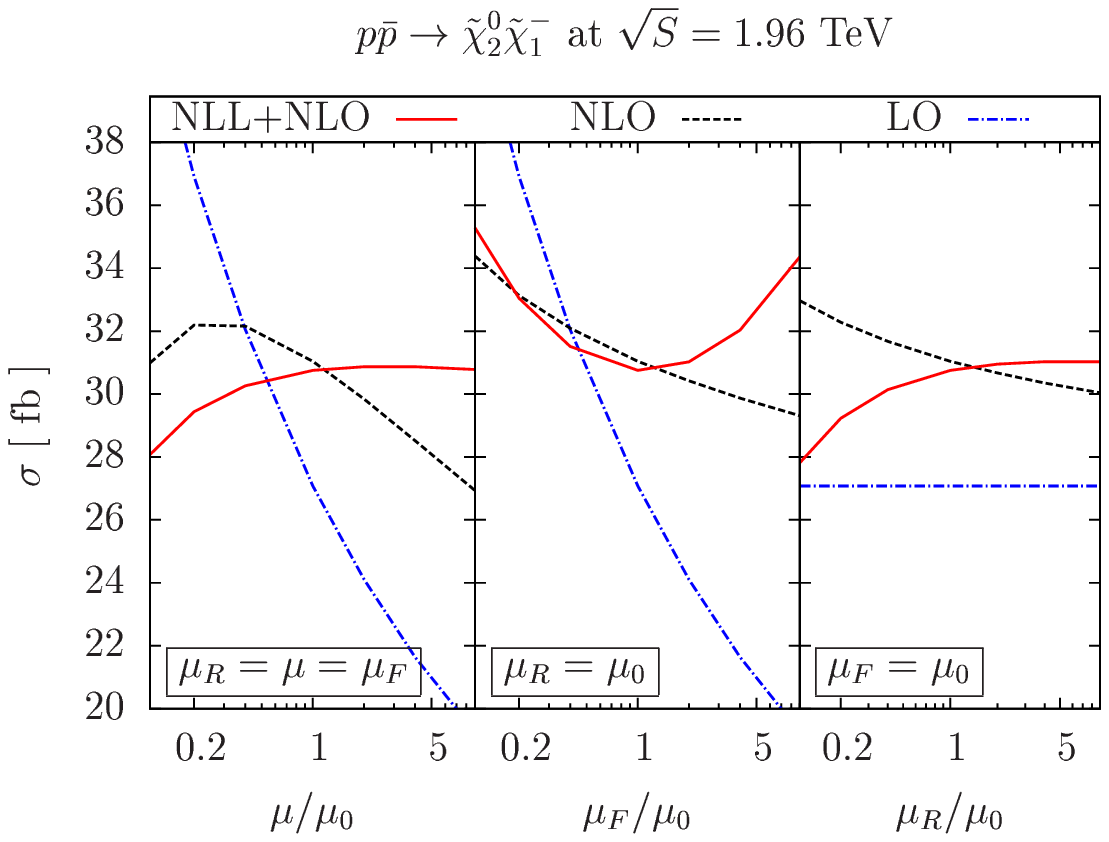}
  \includegraphics[width=.49\textwidth]{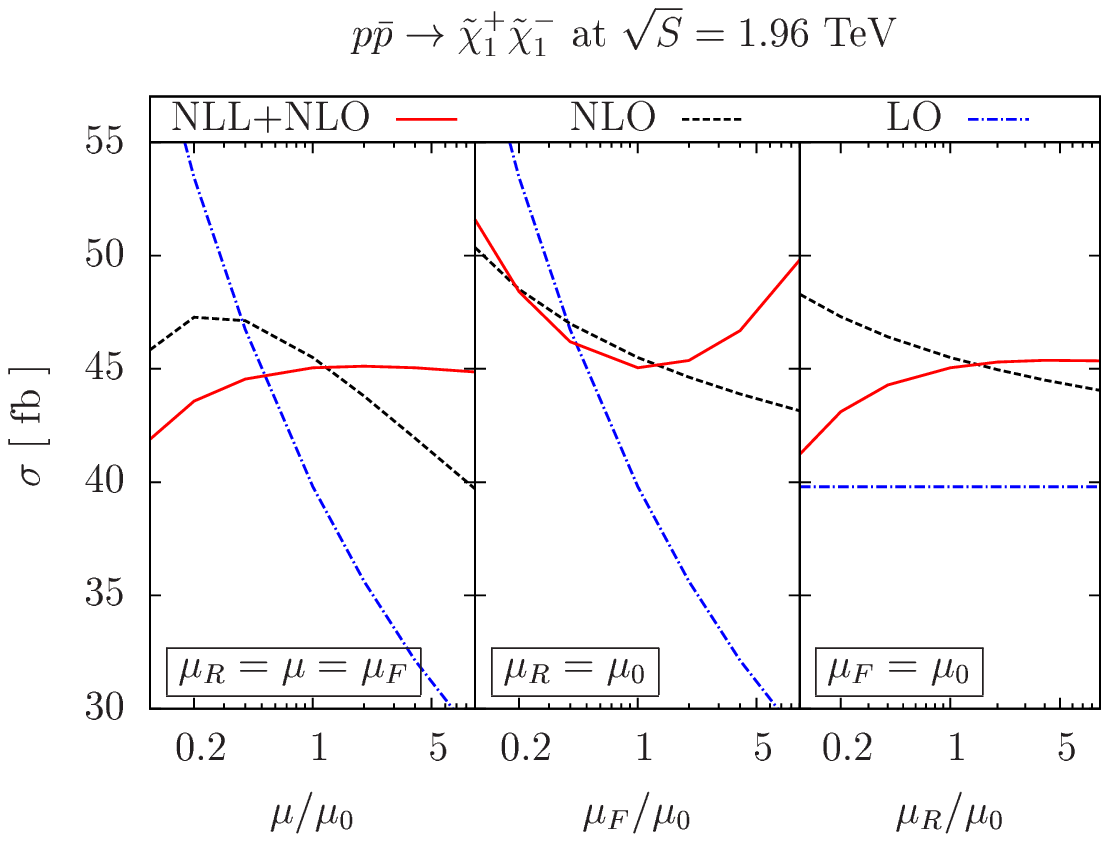}
  \caption{\label{fig:16}Total cross sections for the production of neutralino
 (top left), chargino-neutralino (top right) and chargino pairs
 (bottom) at the Tevatron
 with $\sqrt{S}=1.96$ TeV in the LO (blue, dot-dashed), NLO (black, dashed) and
 NLL+NLO (red, full) approximation.}
\end{figure}

\begin{figure}[t]
  \centering
  \includegraphics[width=.49\textwidth]{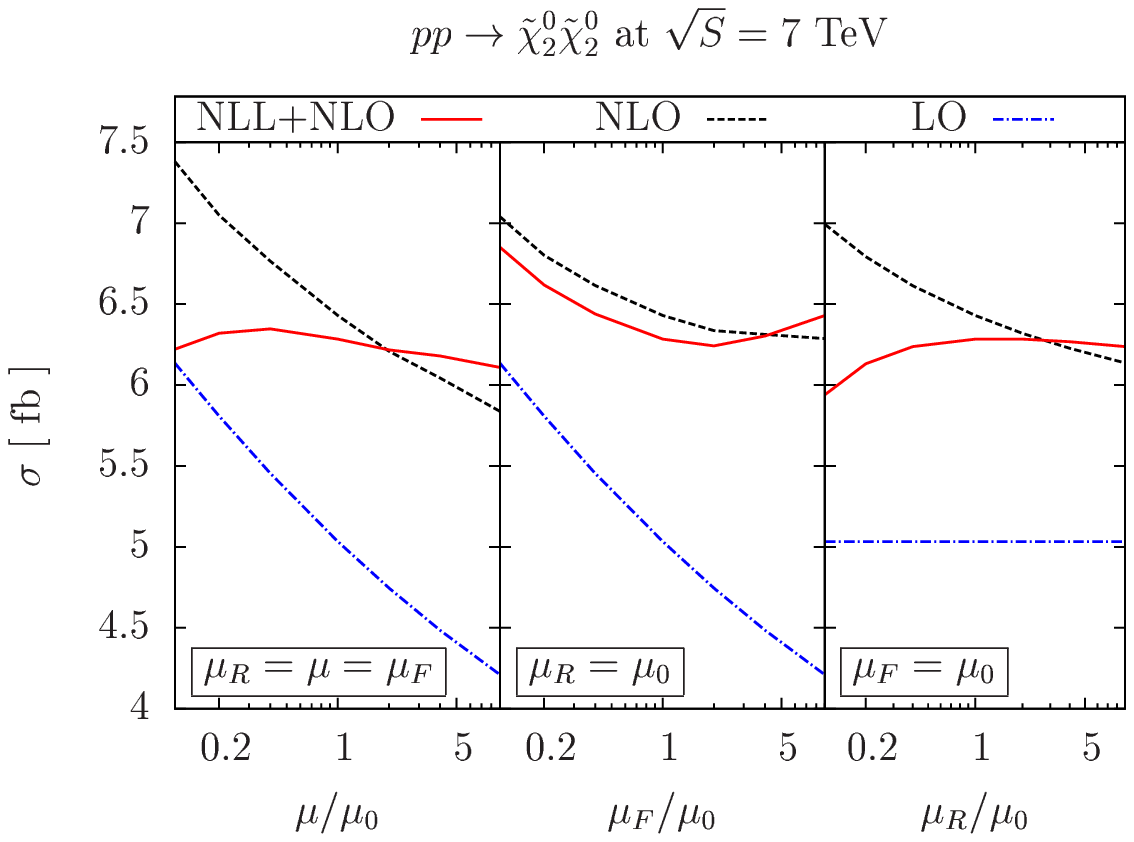}
  \includegraphics[width=.49\textwidth]{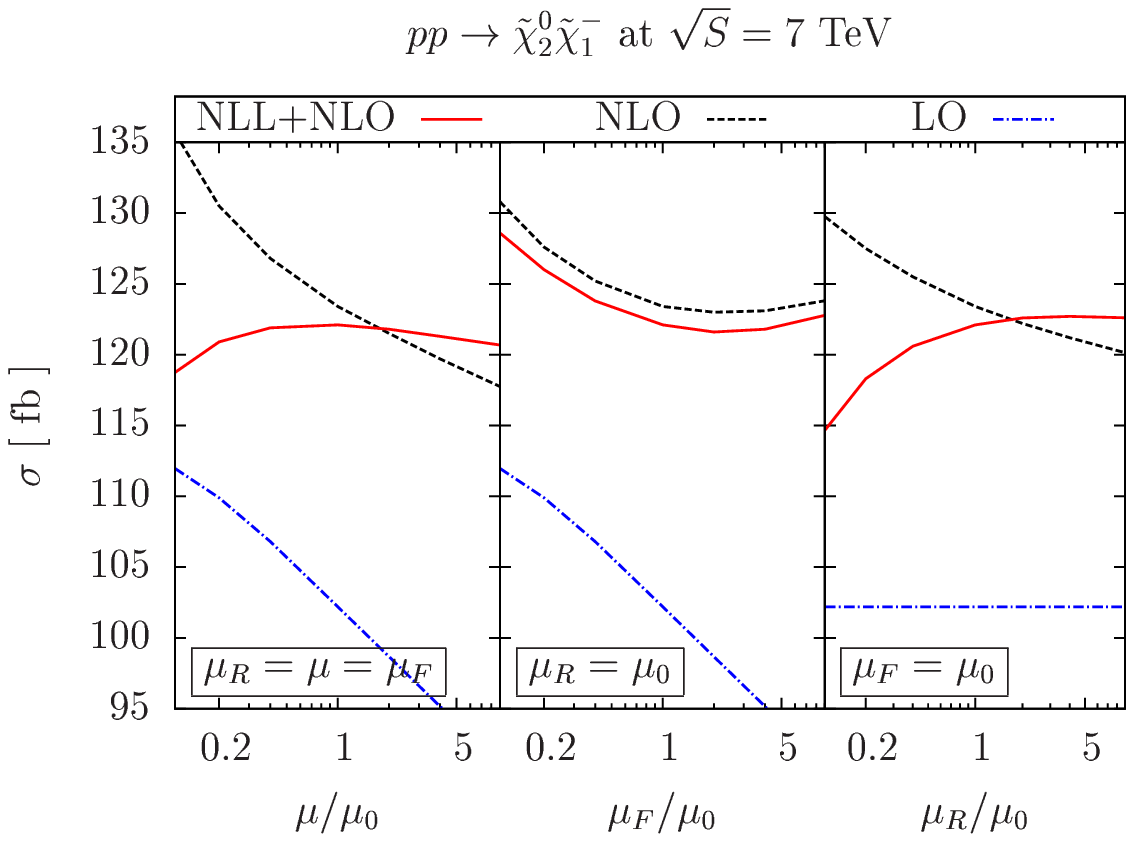}
  \includegraphics[width=.49\textwidth]{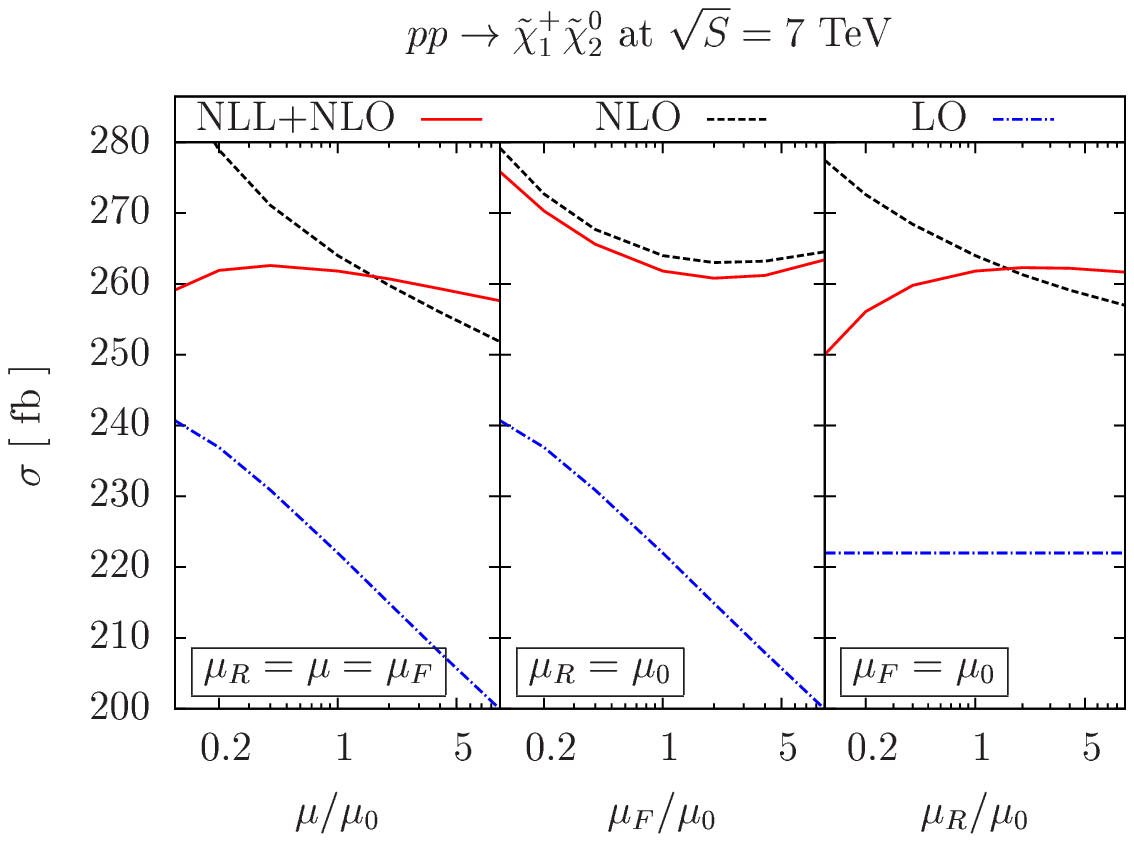}
  \includegraphics[width=.49\textwidth]{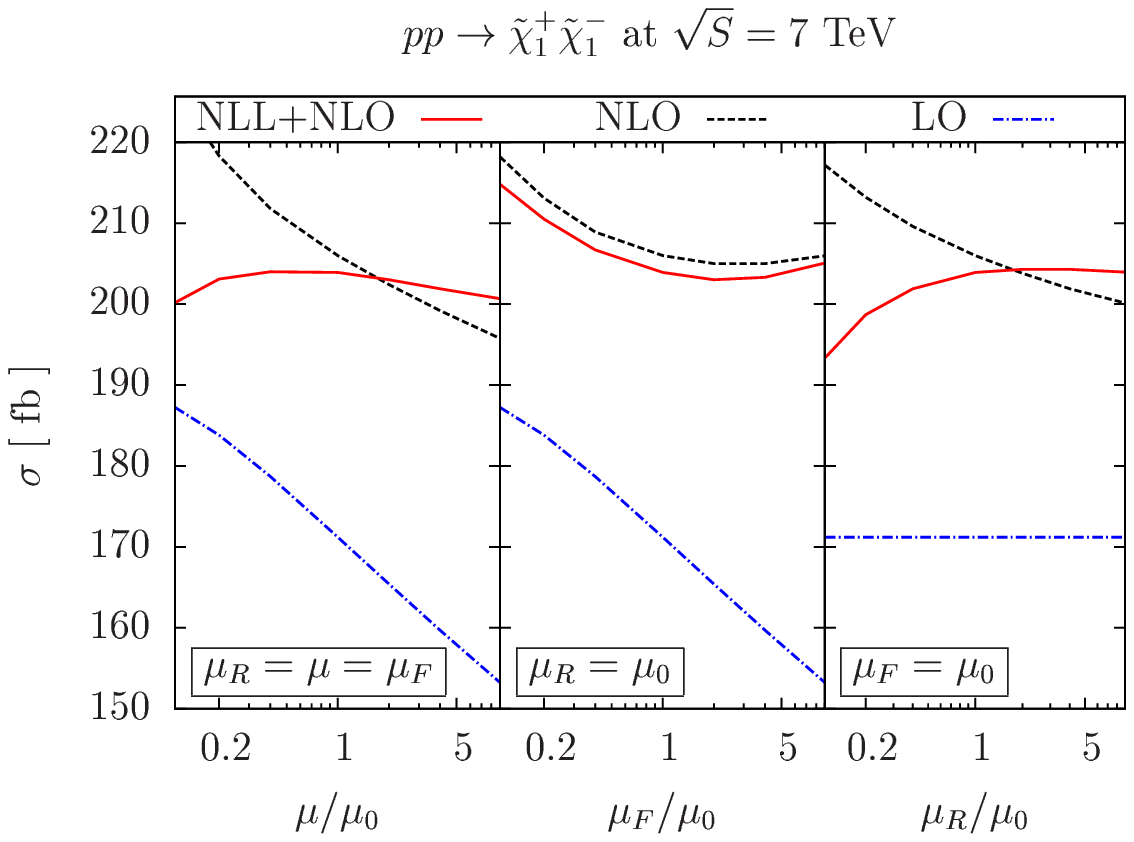}
  \caption{\label{fig:17}Same as Fig.\ \ref{fig:16} for the LHC with its current
 center-of-mass energy of $\sqrt{S}=7$ TeV.}
\end{figure}

The stability of the perturbative series and its reorganization is traditionally
checked by varying the factorization and renormalization scales $\mu_F$ and 
$\mu_R$ about a central value $\mu_0$. We therefore present now the total cross
sections for the production of light gaugino pairs at the Tevatron (Fig.\
\ref{fig:16}) and
at the LHC with $\sqrt{S}=7$ TeV (Fig.\ \ref{fig:17})
as a
%
%
%
function of the ratio $\mu_{F,R}/\mu_0$, where the central scale $\mu_0$ is now
chosen to be the average mass of the produced gaugino pair. The LO prediction
(blue, dot-dashed) of the electroweak processes under consideration is, of course,
independent of the renormalization scale $\mu_R$ (right part of the figures),
whereas the NLO prediction (black, dashed) depends inversely on the logarithm of
$\mu_R$ through the strong coupling $\alpha_s(\mu_R)$. At NLL accuracy (red, full),
the resummed soft corrections attenuate this dependence and introduce a plateau
region, so that the prediction is stabilized. The factorization scale $\mu_F$
(central part of the figures) enters the hadronic cross section already at LO
through the largely logarithmic dependence of the PDFs, which is then attenuated
by the factorization of initial-state singularities at NLO and further at NLL
accuracy. In all cases, the resulting total NLL+NLO prediction is thus much less
dependent on the common scale $\mu_F=\mu_R=\mu$ (left part of the figures) than
the LO and NLO estimates.

In Tab.\ \ref{tab:3} we present the total cross sections for the trilepton
\begin{table}[t]
  \centering
  \caption{\label{tab:3}Total cross sections for the production of $\tilde{\chi}_1
 ^+\tilde{\chi}_2^0$ pairs in the SPS1a' scenario at different hadron colliders
 and center-of-mass energies in the LO, NLO and NLL+NLO approximation, together
 with the corresponding scale and PDF uncertainties.\\}
  \begin{tabular}{|c|c|c|c|}
    \hline Colliders & $\sigma^{\rm NLL+NLO}$ [fb] & $\sigma^{\rm NLO}$ [fb] & $\sigma^{\rm LO}$ [fb]
    \\ \hline \hline $p\bar{p}$($\sqrt{S}=1.96$ TeV)
    & $30.9^{+0.1}_{-0.2}\;^{+1.5}_{-1.9}$ & $31.2^{+0.9}_{-1.2}\;^{+1.5}_{-1.9}$ & $27.2^{+3.6}_{-3.0}$ 
    \\ \hline $pp$($\sqrt{S}=7$ TeV)
    & $263.3^{+0.6}_{-1.3}\;^{+11.4}_{-13.2}$ & $265.5^{+5.0}_{-4.3}\;^{+11.5}_{-13.2}$ & $223.1^{+6.9}_{-7.1}$ 
    \\ \hline $pp$($\sqrt{S}=10$ TeV)
    & $470.7^{+1.4}_{-2.3}\;^{+17.7}_{-19.3}$ & $474.0^{+8.3}_{-6.0}\;^{+17.7}_{-19.4}$ & $387.4^{+2.5}_{-4.3}$ 
    \\ \hline $pp$($\sqrt{S}=14$ TeV)
    & $772.7^{+1.6}_{-3.1}\;^{+25.5}_{-26.7}$ & $777.5^{+11.9}_{-7.4}\;^{+25.5}_{-26.7}$ & $623.7^{+4.7}_{-9.3}$
    \\ \hline
  \end{tabular}
\end{table}
channel in the SPS1a' scenario at the Tevatron ($\sqrt{S}=1.96$ TeV) and LHC
($\sqrt{S}=7$, 10 and 14 TeV). Besides the central values (in fb) at LO, NLO and
NLL+NLO, we also present the scale and PDF uncertainties. The former are
estimated as described above by a common variation of the renormalization and
factorization scales by a factor of two about the average mass of the two
gauginos, the latter through
\bea
 \Delta\sigma_{\rm PDF+}~=~\sqrt{\sum_{i=1}^{22}\le\max\lr\sigma_{+i}-\sigma_0,
 \sigma_{-i}-\sigma_0,0\rr\re^2}&,&
 \Delta\sigma_{\rm PDF-}~=~\sqrt{\sum_{i=1}^{22}\le\max\lr\sigma_0-\sigma_{+i},
 \sigma_0-\sigma_{-i},0\rr\re^2}
\eea
along the 22 eigenvector directions
defined by the CTEQ collaboration. Since these are available only for the NLO
fit CTEQ6.6M, but not for the LO fit CTEQ6.6L1, we do not present a PDF
uncertainty for the LO prediction. Furthermore, the same PDF set enters at NLO
and NLL+NLO, so that the PDF uncertainties for these two predictions coincide.
The most important result is again the considerable reduction of the scale
uncertainty from LO to NLO and then to NLL+NLO. The total cross sections increase
with the available center-of-mass energy due to the higher parton luminosity at
smaller values of $x$. A crude estimate gives
\bea
 \sigma_{pp}&=& \int_{m^2/S}^1\d\tau\,f_{q/p}(x_q)\,f_{\bar{q}/p}(x_{\bar{q}})\,\sigma_{q\bar{q}}
 ~\sim~\int_{m^2/S}^1\d\tau\,\tau^{-1.8}{1\over \tau S},
 ~\sim~\sqrt{S}^{1.6}
\eea
which agrees with the cross sections given in Tab.\ \ref{tab:3} surprisingly well.

In Tab.\ \ref{tab:4} we fix the LHC center-of-mass energy to its design value
\begin{table}[t]
  \centering
  \caption{\label{tab:4}Total cross sections for the production of various
 gaugino pairs in the SPS1a' scenario at the LHC with its design center-of-mass
 energy of $\sqrt{S}=14$ TeV. The central predictions are given at LO, NLO and
 NLL+NLO together with the corresponding scale and PDF uncertainties.\\}
  \begin{tabular}{|c|c|c|c|}
    \hline Gaugino pair& $\sigma^{\rm NLL+NLO}$ [fb]
    & $\sigma^{\rm NLO}$ [fb] & $\sigma^{\rm LO}$ [fb]
    \\ \hline \hline $\tilde{\chi}_2^0 \tilde{\chi}_2^0$
    & $25.1^{+0.3}_{-0.2}\;^{+1.2}_{-0.7}$ & $25.5^{+0.8}_{-0.6}\;^{+1.3}_{-0.7}$ & $19.2^{+0.3}_{-0.4}$ 
    \\ \hline $\tilde{\chi}_1^+ \tilde{\chi}_1^-$
    & $665.8^{+1.0}_{-2.2}\;^{+20.6}_{-20.6}$ & $671.1^{+10.8}_{-6.6}\;^{+20.7}_{-20.6}$ & $533.4^{+3.4}_{-7.3}$ 
    \\ \hline $\tilde{\chi}_2^0 \tilde{\chi}_1^-$
    & $433.3^{+0.6}_{-0.8}\;^{+17.9}_{-16.0}$ & $436.9^{+7.2}_{-3.5}\;^{+17.9}_{-16.1}$ & $348.3^{+2.2}_{-4.9}$ 
    \\ \hline $\tilde{\chi}_1^+ \tilde{\chi}_2^0$
    & $772.7^{+1.6}_{-3.1}\;^{+25.5}_{-26.7}$ & $777.5^{+11.9}_{-7.4}\;^{+25.5}_{-26.7}$ & $623.7^{+4.7}_{-9.3}$
    \\ \hline $\tilde{\chi}_3^0 \tilde{\chi}_4^0$
    & $14.6^{+0.0}_{-0.1}\;^{+0.7}_{-0.7}$ & $14.8^{+0.3}_{-0.3}\;^{+0.7}_{-0.7}$ & $12.1^{+0.5}_{-0.5}$ 
    \\ \hline $\tilde{\chi}_2^+ \tilde{\chi}_2^-$
    & $14.0^{+0.1}_{-0.0}\;^{+0.7}_{-0.6}$ & $14.2^{+0.3}_{-0.3}\;^{+0.7}_{-0.6}$ & $11.7^{+0.5}_{-0.5}$ 
    \\ \hline $\tilde{\chi}_3^0 \tilde{\chi}_2^-$
    & $8.5^{+0.0}_{-0.0}\;^{+0.6}_{-0.5}$ & $8.6^{+0.2}_{-0.2}\;^{+0.6}_{-0.5}$ & $6.9^{+0.3}_{-0.3}$ 
    \\ \hline $\tilde{\chi}_2^+ \tilde{\chi}_3^0$
    & $19.1^{+0.1}_{-0.1}\;^{+0.9}_{-1.0}$ & $19.3^{+0.4}_{-0.4}\;^{+0.9}_{-1.0}$ & $16.0^{+0.7}_{-0.7}$ 
    \\ \hline $\tilde{\chi}_4^0 \tilde{\chi}_2^-$
    & $7.8^{+0.0}_{-0.0}\;^{+0.5}_{-0.5}$ & $7.9^{+0.2}_{-0.2}\;^{+0.5}_{-0.5}$ & $6.4^{+0.3}_{-0.3}$ 
    \\ \hline $\tilde{\chi}_2^+ \tilde{\chi}_4^0$
    & $17.7^{+0.1}_{-0.1}\;^{+0.8}_{-0.9}$ & $17.8^{+0.4}_{-0.3}\;^{+0.8}_{-0.9}$ & $14.9^{+0.7}_{-0.6}$ 
    \\ \hline
  \end{tabular}  
\end{table}
of $\sqrt{S}=14$ TeV and show the total production cross sections for light
and heavy gaugino pairs in LO, NLO and NLL+NLO together with the corresponding
theoretical uncertainties. As it was already mentioned above, the cross section
for the higgsino-like $\nc\nd$ pairs is about as large as the one for
$\tilde{\chi}_2^+\tilde{\chi}_2^-$ pairs, and it is in fact not much smaller than
the one for the considerably lighter gaugino-like $\nb\nb$ pairs. In general,
the heavy gaugino cross sections are, however, significantly smaller than those
for light gauginos.

In Tab.\ \ref{tab:5}, we present finally total cross sections for the trilepton
\begin{table}[t]
  \centering
  \caption{\label{tab:5}Total cross sections for the production of $\tilde{\chi}_1
 ^+\tilde{\chi}_2^0$ pairs at the LHC with its current center-of-mass energy of
 $\sqrt{S}=7$ TeV for different SUSY benchmark points. The central predictions
 are given at LO, NLO and NLL+NLO together with the corresponding scale and PDF
 uncertainties.\\}
  \begin{tabular}{|c|c|c|c|}
    \hline Scenario & $\sigma^{\rm NLL+NLO}$ [fb] & $\sigma^{\rm NLO}$ [fb] & $\sigma^{\rm LO}$ [fb]
    \\ \hline \hline LM1
    & $294.6^{+0.8}_{-1.4}\;^{+12.8}_{-14.5}$ & $297.0^{+5.8}_{-4.8}\;^{+12.8}_{-14.5}$ & $248.2^{+7.1}_{-7.5}$ 
    \\ \hline LM7
    & $538.9^{+2.4}_{-3.5}\;^{+23.9}_{-26.7}$ & $543.8^{+12.8}_{-10.7}\;^{+24.0}_{-26.8}$ & $441.2^{+14.0}_{-14.3}$ 
    \\ \hline LM9
    & $1736.2^{+8.0}_{-12.1}\;^{+68.8}_{-74.3}$ & $1750.3^{+38.8}_{-28.8}\;^{+69.0}_{-74.4}$ & $1374.4^{+8.4}_{-15.7}$ 
    \\ \hline SU2
    & $171.7^{+0.5}_{-0.9}\;^{+8.5}_{-9.8}$ & $173.4^{+4.2}_{-3.9}\;^{+8.5}_{-9.8}$ & $145.0^{+7.4}_{-7.0}$ 
    \\ \hline SU3
    & $116.9^{+0.1}_{-0.4}\;^{+5.6}_{-6.4}$ & $118.0^{+2.2}_{-2.1}\;^{+5.6}_{-6.4}$ & $101.6^{+4.6}_{-4.4}$ 
    \\ \hline SU2+JV
    & $170.4^{+0.2}_{-0.7}\;^{+8.6}_{-9.8}$ & $172.0^{+3.9}_{-3.6}\;^{+8.6}_{-9.9}$ & $145.0^{+7.4}_{-7.0}$ 
    \\ \hline SU3+JV
    & $115.4^{+0.1}_{-0.1}\;^{+5.6}_{-6.4}$ & $116.6^{+1.9}_{-1.8}\;^{+5.6}_{-6.4}$ & $101.6^{+4.6}_{-4.4}$ 
    \\ \hline
  \end{tabular}
\end{table}
channel in our different benchmark scenarios at the LHC with its current
center-of-mass energy of $\sqrt{S}=7$ TeV. Since the masses of $\ca$
and $\nb$ are always rather similar, one expects also similar total cross
sections. This is indeed confirmed by the $\sqrt{S}=7$ TeV results in Tab.\
\ref{tab:3} and the numbers in Tab.\ \ref{tab:5} with the notable exceptions of
LM7 and LM9, where the cross section is about a factor of two and one order of
magnitude larger than for the other points, respectively. This is partly due
to the lower gaugino masses at LM9 and partly to the much heavier squark masses,
which suppress the
$t$- and $u$-channels and thus their destructive interference with the
$s$-channel amplitudes. The additional jet veto (JV), i.e.\ the rejection of
events containing jets with transverse momentum $p_T>20$ GeV, envisaged by the
ATLAS collaboration to suppress the background from top quark pair
production, has obviously no consequences at LO, since gauginos are exclusively
produced at this order. An additional quark or gluon can only be present
at NLO or NLL+NLO, and restricting its transverse momentum to low values
reduces the total cross sections slightly with respect to the unrestricted
predictions. The small reduction of the signal cross section in combination
with a large reduction of the background should therefore indeed lead to a much
better significance.

\section{Conclusions}
\label{sec:5}

In summary, we presented in this paper a complete analysis of threshold
resummation effects on direct light and heavy gaugino pair production at the Tevatron
and at the LHC with its current, intermediate, and design center-of-mass energies.
We confirmed the existing NLO calculation and extended it to include also mixing
effects for intermediate squarks. Soft gluon radiation in the threshold region
was resummed at leading and next-to-leading logarithmic accuracy into a Sudakov
exponential, and the full SUSY-QCD corrections were retained in the finite
coefficient function. This allowed us to correctly match the resummed cross
section at NLL accuracy to the NLO perturbative cross section. Universal subleading
logarithms coming from the splitting of initial partons were resummed in full
matrix form, i.e.\ including also non-diagonal splittings.
We found that resummation
increased the invariant mass spectra and total cross sections only slightly,
but stabilized the reorganized perturbative series considerably with respect
to the fixed-order calculation. For future reference, we presented total cross
sections with the corresponding theoretical errors from scale and PDF variations
at LO, NLO and NLL+NLO in tabular form for several commonly used SUSY benchmark
points, light and heavy gaugino pairs, and various hadron collider types and
energies.

\newpage

\appendix
\section{Quark and squark couplings to weak gauge bosons and gauginos}
\label{sec:a}

The coupling strengths of the electroweak gauge bosons $Z$ and $W$ to left- and
right-handed (s)quarks $q$ ($\sq$) with weak isospin $T^3_q$ and fractional
electric charge $e_q$ are given by
\bea
    \{ L_{qqZ}, R_{qqZ} \} &=&\ -\frac{1}{2 c_W} (T^3_q - e_q\,x_W)\ ,\nonumber\\
    \{ L_{\tilde{q}_i \tilde{q}_j Z}, R_{\tilde{q}_i \tilde{q}_j Z} \} &=&\ -\frac{1}{2 c_W} (T^3_q - e_q\,x_W) \{R^{\sq}_{i1}\, R^{\sq\ast}_{j1}, R^{\sq}_{i2}\, R^{\sq\ast}_{j2} \}\ ,\nonumber\\ 
    \{L_{udW}, R_{udW}\} &=&\ \{ -{V_{ud} \over 2 c_W}, 0\}\ ,\nonumber\\ 
    \{L_{\tilde{u}_i \tilde{d}_j W}, R_{\tilde{u}_i \tilde{d}_j W}\} &=&\ \{-\frac{V_{ud}}{2 c_W} R^{\su}_{i1}\, R^{\sd\ast}_{j1},\, 0\}, 
\eea
where $x_W=1-c_W^2=s_W^2=\sin^2\theta_W$ is the squared sine of the electroweak
mixing angle, $R^{\su,\sd}_{ij}$ are the elements of the rotation matrices
diagonalizing the up- and down-type squark mass matrices, and $V_{ud}$ are the
elements of the CKM matrix. 
Their SUSY counterparts, i.e.\ the squark-quark-gaugino couplings, are given by
\bea
  L_{\tilde{d}_j d \tilde{\chi}^0_i} &=&\  \frac{(e_d - T^3_d)\, s_W\, N_{i1} + T^3_q\, c_W\, N_{i2}}{\sqrt{2}\, c_W} R^{\sd\ast}_{j1} 
     + \frac{m_{d}\, N_{i3}}{2\sqrt{2}\, m_W\, \cos\beta} R^{\sd\ast}_{j2}\ ,~ \nonumber\\
  R_{\tilde{d}_j d \tilde{\chi}_i^0} &=&\  -\frac{e_d\, s_W\, N^\ast_{i1}}{\sqrt{2}\, c_W} R^{\sd\ast}_{j2} 
     + \frac{m_{d}\, N^\ast_{i3}}{2\sqrt{2}\, m_W\, \cos\beta}R^{\sd\ast}_{j1}\ ,~\nonumber\\
  L_{\tilde{u}_j u \tilde{\chi}^0_i} &=&\  \frac{(e_u - T^3_u)\, s_W\, N_{i1} + T^3_q\, c_W\, N_{i2}}{\sqrt{2}\, c_W} R^{\su\ast}_{j1} 
     + \frac{m_{u}\, N_{i4}}{2\sqrt{2}\, m_W\, \sin\beta}R^{\su\ast}_{j2}\ ,~\nonumber\\
  R_{\tilde{u}_j u \tilde{\chi}_i^0} &=&\  - \frac{e_u\, s_W\, N_{i1}^\ast}{\sqrt{2} c_W} R^{\su\ast}_{j2} 
     + \frac{m_{u}\, N_{i4}^\ast}{2\sqrt{2}\, m_W\, \sin\beta} R^{\su\ast}_{j1} \ ,~\nonumber\\
  L_{\tilde{d}_j u \tilde{\chi}_i^\pm}&=&\  \frac{1}{2} 
     \bigg[ U_{i1}\, R^{\sd\ast}_{j1} - \frac{m_{d}\, U_{i2}}{\sqrt{2}\, m_W\, \cos\beta} R^{\sd\ast}_{j2} \bigg] V_{u d} \ ,~\nonumber\\
  R_{\tilde{d}_j u \tilde{\chi}_i^\pm} &=&\ -
     \frac{m_{u}\, V_{i2}^\ast\, V_{u d}}{2\sqrt{2}\, m_W\, \sin\beta} R^{\sd\ast}_{j1} \ ,~\nonumber\\ 
  L_{\tilde{u}_j d \tilde{\chi}_i^\pm}&=&\  \frac{1}{2}
     \bigg[ V_{i1}\, R^{\su\ast}_{j1} - \frac{m_{u}\, V_{i2}}{\sqrt{2}\, m_W\, \sin\beta}R^{\su\ast}_{j2}\bigg] V_{u d}^\ast \ ,~\nonumber\\ 
  R_{\tilde{u}_j d \tilde{\chi}_i^\pm} &=&\  -
     \frac{m_{d}\, U^\ast_{i2} V^\ast_{u d}}{2\sqrt{2}\, m_W\, \cos\beta} R^{\su\ast}_{j1}~,
\eea
where the matrices $N$, $U$ and $V$ diagonalize the neutral and charged
gaugino-higgsino mass matrices, $m_{u,d}$ are the up- and down-type quark masses,
and $m_W$ is the mass of the $W$-boson. All other couplings
vanish due to electromagnetic charge conservation  \cite{Nilles:1983ge}. 

\acknowledgments

This work has been supported by a Ph.D.\ fellowship of the French ministry
for education and research and by the Theory-LHC-France initiative of the
CNRS/IN2P3.


\end{document}